\documentclass[english]{article}

\usepackage[T1]{fontenc}
\usepackage[latin9]{inputenc}
\usepackage{geometry}
\usepackage{booktabs}
\usepackage{units}
\usepackage{mathrsfs}
\usepackage{amsmath}
\usepackage{amssymb}
\usepackage{graphicx}
\usepackage[authoryear]{natbib}

\makeatletter

\providecommand{\tabularnewline}{\\}

\newcommand{\lyxaddress}[1]{
	\par {\raggedright #1
	\vspace{1.4em}
	\noindent\par}
}

\makeatother

\usepackage{babel}
\begin{document}
\title{Deep water gravity wave triad resonances on uniform flow}
\author{David M. Kouskoulas$^{1,}$\footnote{dkouskoulas@mail.tau.ac.il} and Yaron Toledo$^{1,}$\footnote{toledo@tauex.tau.ac.il}}
\maketitle

\lyxaddress{$^{1}$Marine Engineering and Physics Laboratory, School of Mechanical
Engineering, Tel Aviv University, Tel Aviv 6997801, Israel}
\begin{abstract}
Triad resonances for gravity waves propagating in opposite direction
with respect to uniform current are introduced. They are produced
by multivalued and anisotropic dispersion and occur even in deep water.
In contrast, existing literature suggests non-degenerate  deep water
triads require inhomogeneity or capillary effects. In this sense,
the new resonances are a first of their kind. Analytical conditions
for the existence of resonance may be reduced to universal constants.
Solution of a three-wave interaction model, adapted to the wave-current
problem, shows regimes wherein the dominant direction, speed and strength
of resonant energy transfer differs between wavenumber scales. This
suggests uniform current is a fundamental source of spatial inhomogeneity.
The new resonances significantly modify the picture of nonlinear interactions
in wave-current fields. Since the resonances are quadratic, they will
also dominate well known quartet interactions.
\end{abstract}

\section{Introduction}

Water waves often propagate on flows. The transformation of the single-valued
water wave dispersion relation to a multi-valued and anisotropic form
when accounting for mean motion is well known \citep[cf.][]{kitaigordskii1975phillips,Peregrine76}.
Nonlinear interactions involving the ``small'' wavenumber solutions
have been shown to produce new spatiotemporal features in wave fields
\citep{kouskoulas2017effects}. Similarly, for ships moving with constant
velocity, multi-valued dispersion solutions are of known importance
to heave and pitch ship motions \citep{newman1980unified} and wave
diffraction by slender bodies \citep{sclavounos1981interaction,sclavounos1981diffraction}.
Wave-current experiments have demonstrated the effects of current
direction on Bragg resonance \citep{Rey2002,Magne2005}. Current inhomogeneities
have also been shown to produce resonance by means of wave trapping
\citep{shrira2014nonlinear} and vorticity waves \citep{Drivas2015}.

Resonances are understood to be a dominant mechanism of energy transfer
\citep{hasselmann1962non}. Accordingly, current effects on resonance
are of significant importance. Resonance conditions occur when dispersion
permits phase matching between linear and nonlinear terms. This phase
matching results in growth of nonlinear terms \citep{phillips1960dynamics}.
If the interaction time is sufficiently long, higher order terms may
grow to first-order magnitude \citep{benney1962non}. Resonant energy
exchange has been shown for gravity-capillary waves \citep{mcgoldrick1965resonant},
surface and internal waves \citep{alam2010oblique}, and antiparallel
acoustic and gravity waves \citep{kadri2013generation,kadri2016resonant}.
Resonances, in 2DH problem, have also been suggested near the cusp
line of Kelvin ship waves \citep{newman1970third}. In deep water,
degenerate triad resonances have been discussed \citep{longuet1962resonant}.
However, existing literature suggests there is no non-degenerate resonance
for  deep water gravity waves without inhomogeneity (i.e. current
shear) or capillary effects. In contrast, it will be shown that nontrivial
 gravity wave triads exist for uniform flows in intermediate and even
deep water. 

The existence of nontrivial deep water gravity wave triad resonance
on uniform current is demonstrated. They occur between gravity waves
propagating in opposite directions with respect to current even in
deep water. In $\mathsection$\ref{sec Linear formulation}, the linear
wave-current dispersion relation is reviewed. It is shown to be multi-valued
and anisotropic. In $\mathsection$\ref{sec: existence of res}, a
geometric approach demonstrates the existence of resonance. In $\mathsection$\ref{sec: 4 Res in lab},
resonance conditions are found for a laboratory viewer. They exist
among, but not between, two wave mode types. If the ratio between
initial harmonics is known, conditions reduce to universal constants.
In order to demonstrate consistency between inertial viewers, resonance
conditions for an observer moving with the current velocity is derived
in $\mathsection$\ref{sec: 5 Res in comoving}. Amplitude coupling
and energy exchange is shown in $\mathsection$\ref{sec: Amplitude coupling and energy exchange}.
Resonant behavior differs between wavenumber scales. A conclusion
and discussion is given in $\mathsection$\ref{sec:Conclusion}. The
new resonances preclude more complex 3D-effects and their weak nonlinearity
would dominate known quartet interactions. Accordingly, results suggest
multivalued and anisotropic dispersion may profoundly modify the dominant
energy transfer mechanisms in wave-current fields.

\subsection*{Preliminary remarks on Galilean invariance\label{subsec:Preliminary-remarks-on}}

Since the functional form of the single-valued plane wave dispersion
relation does not permit for a nontrivial deep water triad resonance,
it is perhaps tempting to assume the following results imply a violation
of Galilean invariance. However, resonance conditions are found by
satisfying the basic interaction condition and the relevant dispersion
relations for each wave. Galilean invariance does not guarantee invariance
of the functional form of the latter.\footnote{The strict definition of invariance of dispersion for waves in moving
media has been discussed by \citet{censor1980dispersion,censor1998electrodynamics,mccall2007relativity}.} This is known in the transformation of the single-valued dispersion
relation to a multi-valued form\footnote{The single-valued disperson relation form referred to is $\sigma=\sqrt{gk\tanh kh}$.
The multi-valued form is $\omega=\pm kU+\sqrt{gk\tanh kh}$.} in the presence of mean motion \citep[cf.][]{kitaigordskii1975phillips,Peregrine76,newman1980unified,sclavounos1981interaction,sclavounos1981diffraction,peregrine1983interaction}.
If the Galilean transformation guaranteed invariance of the dispersion
relation's functional form, one of these would be invalid. In fact,
the single-valued case is just a degenerate case in the mathematical
limit of zero current.

On account of the (single-valued) intrinsic frequency dispersion function
being only a degenerate case of the multi-valued form, caution must
be taken when superposing intrinsic frequency representations. The
intrinsic frequencies of opposing plane waves on current are described
by different $locally$ comoving observers \citep[definition in][]{bretherton1968wavetrains}:
one for waves following, and one for waves opposing current\footnote{The two comoving observers are found through application of a Galilean
transformation in different directions with respect to current, which
is defined by the laboratory observer.}. For a consistent description of the system, one must commit to a
single observer. If one chooses the comoving observer for a wave following
current, all waves following the current will not ``feel'' the mean
motion; their dispersion relation will be the single-valued degenerate
case. At the same time, since the system is not a rigid body, waves
opposing the current will ``feel'' a mean motion relative to the
observer. Hence, their dispersion will require the multi-valued form.
The correct form of dispersion, for each direction of wave motion,
is essential in determining resonance conditions. If the dispersion
relations of all waves composing a resonance condition are derived
from the same governing equations, Galilean invariance is not violated.

\section{Linear formulation \label{sec Linear formulation}}

The linear wave-current dispersion relation is derived using a velocity
potential approach. It is shown to be multivalued and anisotropic. 

\subsection{Linear wave-current dispersion\label{subsec:disp relation}}

Assume wave propagation with a uniform current in water of constant
depth $h$. Irrotational and inviscid flow is assumed. $x$ and $z$
are horizontal and vertical coordinates respectively. $\phi\left(x,z,t\right)$
is a velocity potential and $\eta\left(x,t\right)$ is the surface
elevation. The boundary value problem is given by \citep[cf.][]{whitham1962mass}

\begin{eqnarray}
\nabla^{2}\phi & = & 0,\qquad-h<z<\eta,\label{eq: Laplace}\\
-\frac{\partial\phi}{\partial z} & = & 0,\qquad z=-h,\label{eq: BBC}\\
\frac{\partial\phi}{\partial z}-\frac{\partial\eta}{\partial t}-\frac{\partial\phi}{\partial x}\frac{\partial\eta}{\partial x} & = & 0,\qquad z=\eta,\label{eq: KFSBC}\\
\frac{\partial\phi}{\partial t}+\frac{1}{2}\left(\nabla\phi\right)^{2}+g\eta+Q(t) & = & 0,\qquad z=\eta.\label{eq: DFSBC}
\end{eqnarray}
Define flow quantities as power series in the small parameter $\varepsilon=ka$. 

\begin{eqnarray}
\phi & = & Ux+\varepsilon\tilde{\phi},\label{eq: fi def}\\
\eta & = & \varepsilon\tilde{\eta}.\label{eq: eta def}
\end{eqnarray}
The current velocity $U$ has been introduced in (\ref{eq: fi def}).
Substituting (\ref{eq: fi def}) and (\ref{eq: eta def}) into Taylor
series expansions of (\ref{eq: KFSBC}) and (\ref{eq: DFSBC}) yields

\begin{eqnarray}
\frac{\partial\tilde{\phi}}{\partial z}-\frac{\partial\tilde{\eta}}{\partial t}-\left(U+\frac{\partial\tilde{\phi}}{\partial x}\right)\frac{\partial\tilde{\eta}}{\partial x}+\tilde{\eta}\frac{\partial^{2}\tilde{\phi}}{\partial z^{2}} & = & 0,\qquad z=0,\label{eq: TAYLOR KFSBC Taylored}\\
\frac{\partial\tilde{\phi}}{\partial t}+U\frac{\partial\tilde{\phi}}{\partial x}+\frac{1}{2}\left(\frac{\partial\tilde{\phi}}{\partial x}\right)^{2}+g\tilde{\eta}+\tilde{\eta}\frac{\partial^{2}\tilde{\phi}}{\partial z\partial t}+U\tilde{\eta}\frac{\partial^{2}\tilde{\phi}}{\partial x\partial z} & = & 0,\qquad z=0.\label{eq: Taylor DFSBC Taylored}
\end{eqnarray}
Eliminating $\tilde{\eta}$ allows one to combine (\ref{eq: TAYLOR KFSBC Taylored})
and (\ref{eq: Taylor DFSBC Taylored}) into a single equation of the
form

\begin{eqnarray}
\mathscr{L}\left(\tilde{\phi}\right)+\mathscr{P}^{2}\left(\tilde{\phi},\tilde{\phi}\right) & = & 0,\qquad z=0.\label{eq: Operator combined form}
\end{eqnarray}
$\mathscr{L}\left(\cdot\right)$ and $\mathscr{P}^{2}\left(\cdot\right)$
are linear and quadratic nonlinear operators respectively. Quadratic
nonlinear terms are neglected in the linear approximation. The linear
operator is

\begin{eqnarray}
\mathscr{L}\left(\tilde{\phi}\right) & = & \frac{\partial^{2}\tilde{\phi}}{\partial t^{2}}+g\frac{\partial\tilde{\phi}}{\partial z}+2U\frac{\partial^{2}\tilde{\phi}}{\partial x\partial t}+U^{2}\frac{\partial^{2}\tilde{\phi}}{\partial x^{2}},\qquad z=0.\label{eq:linear operator-1}
\end{eqnarray}
The nonlinear terms who compose $\mathscr{P}^{2}\left(\cdot\right)$
are shown in Appendix \ref{sec:Nonlinear-terms-secularitiy}. A linear
deep water solution for $\mathscr{L}\left(\tilde{\phi}\right)=0$
is well known,

\begin{eqnarray}
\tilde{\phi} & = & a\left(c-U\right)e^{\left|k\right|z}\sin\left(kx-\omega t\right).\label{eq: VP}
\end{eqnarray}
$a$ is the amplitude, $c$ is the wave celerity, $k$ is the wavenumber
and $\omega$ is the absolute frequency. Equations (\ref{eq: VP})
and (\ref{eq:linear operator-1}) yield the linear wave-current dispersion
relation,

\begin{equation}
\omega-kU=\pm\sigma=\pm\sqrt{g\left|k\right|}.\label{eq: wc dispersion-1}
\end{equation}
$\sigma$ is the intrinsic frequency. Equation (\ref{eq: wc dispersion-1})
is solved graphically by plotting $y=\omega-kU$ and $\sigma=\pm\sqrt{g\left|k\right|}$.
\begin{figure}
\begin{centering}
\includegraphics[scale=0.4]{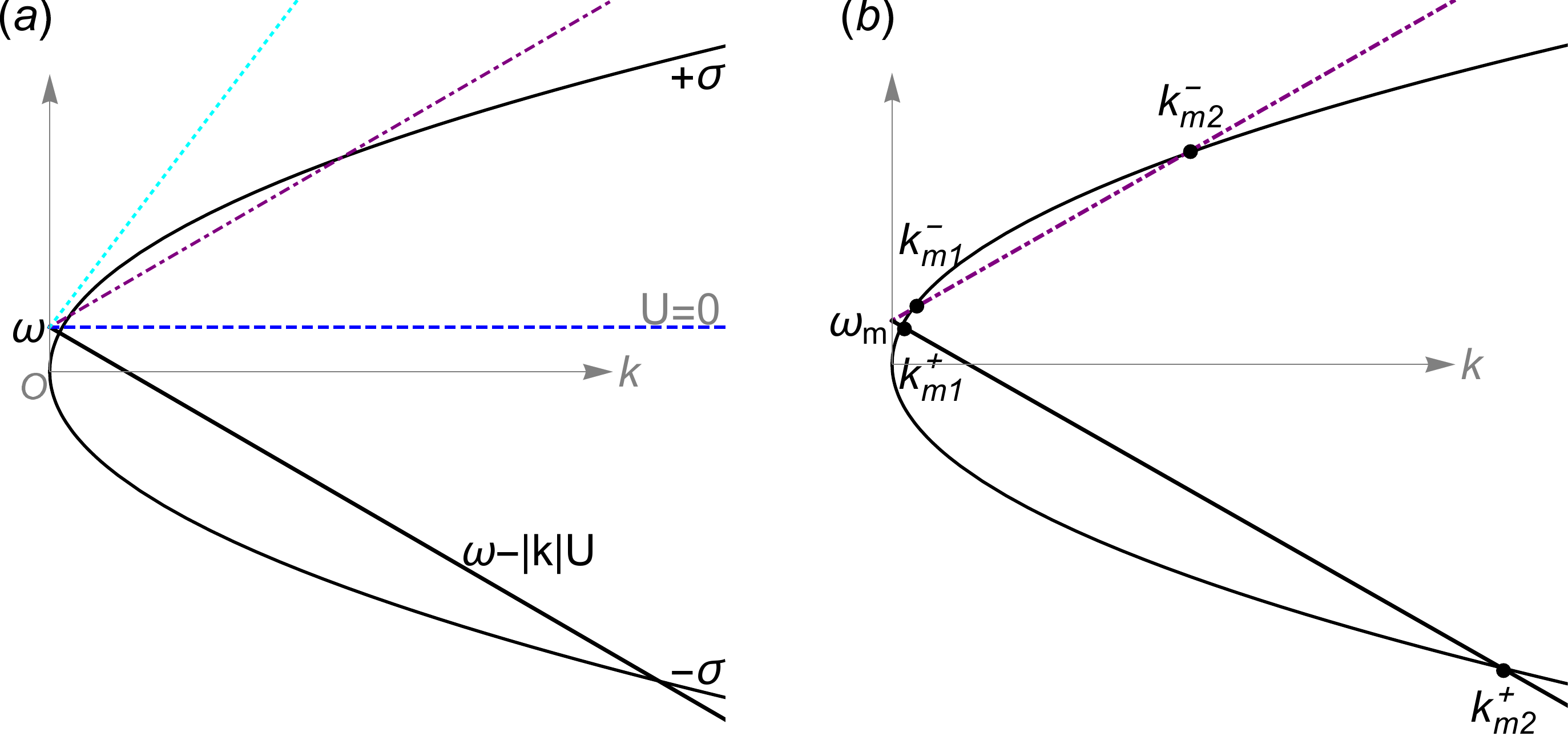}
\par\end{centering}
\centering{}\caption{{\small{}Graphical solution of wave-current dispersion relation. Intersections
of $\mathrm{y=\omega-\mathit{k}\mathit{U}}$ with $\mathrm{\mathit{\sigma}=\pm\left(gk\right)}^{\nicefrac{1}{2}}$
give wavenumber solutions. $+U$ and $-U$ correspond to wave motion
following and opposing current respectively. (a) Solution branches
for various $U$. A single branch may give two; black and purple dot-dashed
lines, one; blue dashed line; or zero; cyan dotted line, solutions.
(b) Wave motion following and opposing current (antiparallel wave
motion) is given by two solution branches and may yield up to four
nontrivial wavenumbers for a single frequency. \label{Dispersion relation graphical solution}}}
\end{figure}
 Wavenumber solutions are given by the points of intersection (see
fig. \ref{Dispersion relation graphical solution}a). ``Small''
and ``large'' wavenumber solutions, for each direction of $U$,
are referred to as type I and type II modes respectively. These two
pairs of wavenumbers are propagating in opposite directions with respect
to current. Moreover, their magnitudes differ for following $\left(U>0\right)$
and opposing $\left(U<0\right)$ current. Thus, wave motion in two
directions with respect to current may yield a total of four nontrivial
wavenumber solutions (see fig. \ref{fig: graphical_solution}b).

\subsection{Analytical solution for laboratory viewer \label{subsec: analytical disp}}

The laboratory viewer sees the fluid move with velocity $U$. The
relevant dispersion relation, (\ref{eq: wc dispersion-1}), may be
written as

\begin{equation}
D\left(k,\omega,U\right)=\omega=kU+\sqrt{g\left|k\right|}.\label{eq: wc dispersion (D form)}
\end{equation}
One may assume only positive wavenumbers. In this convention, waves
propagating in opposite directions will both have positive wavenumbers,
but will be functions of opposite signs of $U$.\footnote{An alternative convention, which defines antiparallel nontrivial wave
modes by oppositely signed wavenumbers, is also possible. The two
conventions and their equivalence are discussed in Appendix \ref{sec:Positive-and-negative wavenumbers}.} In polynomial form (\ref{eq: wc dispersion (D form)}) is

\begin{equation}
\mathscr{D}\left(k,\omega,U\right)=U^{2}k^{2}+\left(-2U\omega-g\right)k+\omega^{2}=0.\label{eq: wc dispersion moving polynomial}
\end{equation}
Since a positive wavenumber convention is used, the absolute value
is dropped but still implied. As was shown graphically in $\mathsection$\ref{sec Linear formulation},
\ref{eq: wc dispersion moving polynomial} is multi-valued. Moreover,
the magnitude of these multiple solutions differs for a change in
sign of $U$. The four nontrivial solutions for (\ref{eq: wc dispersion moving polynomial})
may be written most succinctly in analytical form as
\begin{eqnarray}
k_{mn}^{\pm} & = & \frac{g\pm2U\omega_{m}+\left(-1\right)^{n}\sqrt{g\left(g\pm4U\omega_{m}\right)}}{2U^{2}},\qquad n=1,2.\label{eq: 4 wavenumbers dimensional}
\end{eqnarray}
Wherein, the $\pm$ fixes the direction of wave propagation relative
to current. $-U$ denotes wave propagation against current. $+U$
denotes wave propagation with current. Using this convention, waves
in both directions of current will have positive wavenumbers. $n=1$
and $n=2$ correspond to ``small'' (type I) and ``large'' (type
II) wavenumber solutions respectively. Type II modes have no analogy
for plane wave dispersion in a \textit{locally} comoving frame \citep{peregrine1983interaction}.
The four non-trivial wave mode solutions from (\ref{eq: wc dispersion moving polynomial})
are listed explicitly in Table (\ref{tab:k solutions}).
\begin{table}
\centering{}%
\begin{tabular}{ccccc}
\toprule 
\multicolumn{5}{c}{$\mathscr{D}\left(k,\omega,U\right)$}\tabularnewline
\midrule 
 & $n$ & $U$ & $k_{mn}^{\pm}$ & mode type\tabularnewline
\midrule
\midrule 
$k_{m1}$ & 1 & $+$ & $\frac{g+2U\omega_{m}-\sqrt{g\left(g+4U\omega_{m}\right)}}{2U^{2}}$ & I\tabularnewline
\midrule 
$k_{m2}$ & 2 & $+$ & $\frac{g+2U\omega_{m}+\sqrt{g\left(g+4U\omega_{m}\right)}}{2U^{2}}$ & II\tabularnewline
\midrule 
$k_{m1}^{-}$ & 1 & $-$ & $\frac{g-2U\omega_{m}-\sqrt{g\left(g-4U\omega_{m}\right)}}{2U^{2}}$ & I\tabularnewline
\midrule 
$k_{m2}^{-}$ & 2 & $-$ & $\frac{g-2U\omega_{m}+\sqrt{g\left(g-4U\omega_{m}\right)}}{2U^{2}}$ & II\tabularnewline
\bottomrule
\end{tabular}\caption{Wavenumber solutions for $\mathscr{D}\left(k,\omega,U\right)$ using
a positive wavenumber convention. Laboratory viewer, $\mathscr{D}$,
sees four non-trivial positive wavenumber solutions. $n=1$ and $n=2$
are ``small'' and ``large'' wavenumber solutions (``short''
and ``long'' wave solutions). Waves opposing current are defined
by $-U$, waves following current are given by $U$.\label{tab:k solutions}}
\end{table}

\section{Geometric demonstration of resonance \label{sec: existence of res}}

$Existence$ of resonance conditions for water waves on uniform current
is demonstrated geometrically using the description for a laboratory
observer (absolute frequencies). Resonance occurs from a ``tuning''
of the wave field by dispersion. When a system is in resonance, phases
of linear and nonlinear components match. This causes secular growth
of nonlinear terms \citep{phillips1960dynamics}. The triad resonance
condition is \citep[cf. ][]{hammack1993resonant}

\begin{eqnarray}
\omega_{a}+\omega_{b}-\omega_{c} & = & 0,\label{eq: omega clsoure}\\
k_{a}+k_{b}-k_{c} & = & 0.\label{eq: k closure}
\end{eqnarray}
A resonance $exists$ if (\ref{eq: omega clsoure}) and (\ref{eq: k closure})
can be satisfied along with the dispersion relation. Deriving resonance
conditions is often algebraically tedious. However, $existence$ of
resonance may be easily and unambiguously demonstrated graphically.
The graphical approach involves finding intersections between solution
curves of (\ref{eq: omega clsoure}) and (\ref{eq: k closure}) when
constrained by the dispersion relation. The approach has been used
to demonstrate resonances involving internal waves \citep{ball1964energy},
gravity-capillary waves \citep{simmons1969variational}, and vorticity
waves \citep{Drivas2015}. 

The procedure is as follows: first, plot the solution curve of a dispersion
relation $D_{1}\left(k,\omega\right)$ with its origin at $\mathrm{O}_{1}=\left(0,0\right)$.
Second, plot the solution curve again with its origin at $\mathrm{O}_{2}=\left(k_{a},\omega_{a}\right)$,
wherein $\left(k_{a},\omega_{a}\right)$ is a solution of $D_{1}$
(i.e. $D_{2}\left(k_{a},\omega_{a}\right)$ wherein $\left.\left(k_{a},\omega_{a}\right)\in\mathbb{R}^{+}\right|D_{1}\left(k,\omega\right)=0$).
If there is a point of intersection between $D_{1}$ and $D_{2}$,
it represents a solution to the vector equation

\begin{eqnarray}
\left\{ \begin{array}{c}
\omega_{a}\\
k_{a}
\end{array}\right\} +\left\{ \begin{array}{c}
\omega_{b}\\
k_{b}
\end{array}\right\}  & = & \left\{ \begin{array}{c}
\omega_{c}\\
k_{c}
\end{array}\right\} .\label{eq: vector equation}
\end{eqnarray}
Since this solution falls on both solution curves of the dispersion
relation, it is equivalent to solving the algebraic system given by
(\ref{eq: omega clsoure}) and (\ref{eq: k closure}) along with the
constraint of dispersion. Accordingly, it represents the $existence$
of a triad resonance. 

\begin{figure}
\centering{}\includegraphics[scale=0.4]{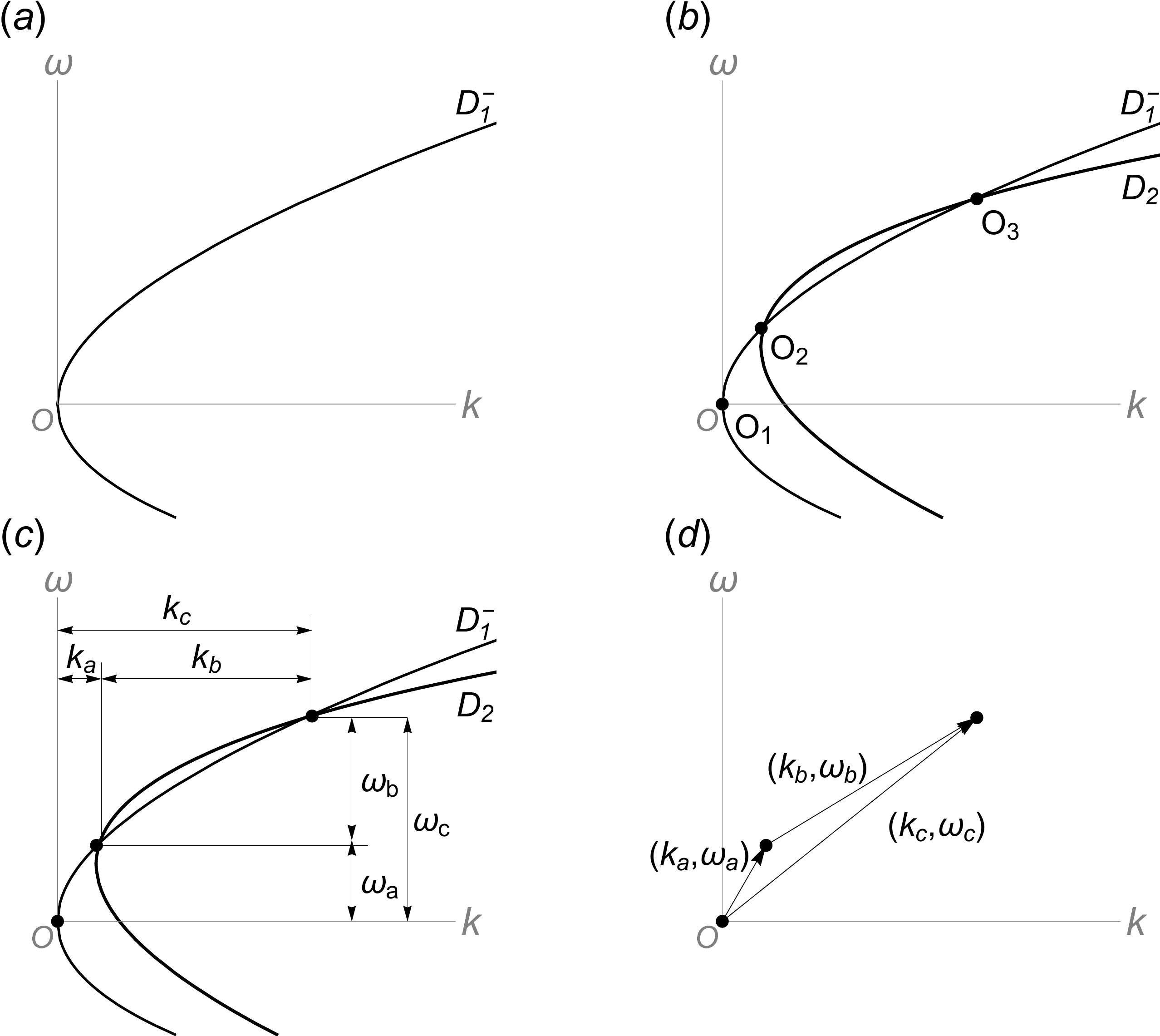}\caption{$Existence$ of resonance for antiparallel gravity waves on uniform
current is demonstrated graphically. (a) First, plot $D_{1}\left(k,\omega,-U\right)=\omega=-kU+\sqrt{g\left|k\right|}$
with origin at $\mathrm{O}_{1}=\left(k,\omega\right)=\left(0,0\right)$.
(b) Second, plot $D_{2}\left(k,\omega,U\right)=\omega=kU+\sqrt{g\left|k\right|}$
with origin at $\mathrm{O}_{2}=\left(k_{a},\omega_{a}\right)$, wherein
$\left(k_{a},\omega_{a}\right)$ is a solution of $D_{1}\left(k,\omega,-U\right)$.
Intersection of solution curves at point $\mathrm{O}_{3}$ represents
the $existence$ of resonance. (c) Resonant wavenumbers and frequencies
are given by $\overline{\mathrm{O}_{1}\mathrm{O}_{2}}$, $\overline{\mathrm{O}_{2}\mathrm{O}_{3}}$
and $\overline{\mathrm{O}_{1}\mathrm{O}_{3}}$. (d) Graphical solution
reduces to vector representation of resonance closure. \label{fig: graphical_solution}}
\end{figure}

In figures \ref{fig: graphical_solution}a-\ref{fig: graphical_solution}d,
the procedure is applied to the wave-current dispersion relation in
the form given by (\ref{eq: wc dispersion (D form)}).
 In figure \ref{fig: graphical_solution}a the solution curve for
$D_{1}\left(k,\omega,-U\right)$ is plotted. In figure \ref{fig: graphical_solution}b,
the solution curve for $D_{2}\left(k,\omega,U\right)$ is plotted
along $D_{1}$. They are seen to intersect at a point $\mathrm{O}_{3}$.
This demonstrates the $existence$ of a triad resonance for deep water
gravity waves propagating in opposite directions on uniform current.
Resonant wavenumbers and frequencies are given by $\overline{\mathrm{O}_{1}\mathrm{O}_{2}}$,
$\overline{\mathrm{O}_{2}\mathrm{O}_{3}}$ and $\overline{\mathrm{O}_{1}\mathrm{O}_{3}}$
(see figure \ref{fig: graphical_solution}c). The graphical solution
simplifies to a vector representation (see figure \ref{fig: graphical_solution}d)
which represents a solution to (\ref{eq: vector equation}). Since
the closure is in terms of absolute frequencies, it represents a resonant
condition for a laboratory observer. The $existence$ of nontrivial
deep water gravity wave triad resonances on uniform current is the
main finding of this paper.

\section{Resonance conditions for a laboratory observer\label{sec: 4 Res in lab}}

\begin{figure}
\begin{centering}
\includegraphics[scale=0.4]{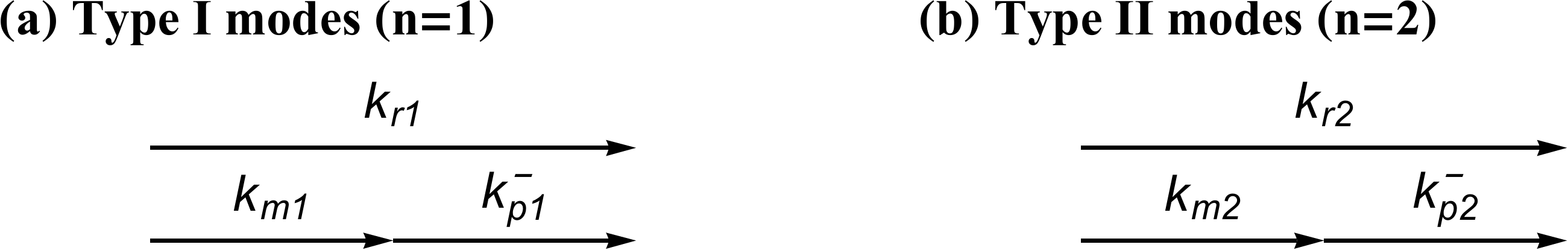}
\par\end{centering}
\centering{}\caption{Resonant wavenumber schematic: antiparallel waves (with respect to
current direction) may resonate on uniform current in deep water.
$-$ superscript denotes wave against current. Subscripts $m,p,r$
denote frequencies which satisfy $\omega_{m}+\omega_{p}=\omega_{r}$.
Second subscript denotes wavenumber type. For resonance, all modes
must be either type I or II. \label{fig:Schematic-of-possible}}
\end{figure}

A laboratory observer sees the fluid move with the current velocity,
$U$, and measures absolute frequencies, $\omega$. Consider three
plane waves: $a_{m},a_{p}^{-}$ and $a_{r}$. $a_{m}$ and $a_{r}$
propagate with current, and $a_{p}^{-}$ propagates against current.
Introduce (\ref{eq: 4 wavenumbers dimensional}) into (\ref{eq: omega clsoure})
and (\ref{eq: k closure}). The triad closure for a laboratory observer
is

\begin{eqnarray}
\omega_{m}+\omega_{p}-\omega_{r} & = & 0,\label{eq: freq closure}\\
k_{mn}^{\pm}+k_{pq}^{\pm}-k_{rs}^{\pm} & = & 0.\label{eq: wavenumber closure}
\end{eqnarray}
The velocity potential is

\begin{eqnarray}
\tilde{\phi} & = & \left\{ b_{mn}e^{k_{mn}z}e^{i\theta_{mn}}+c.c.\right\} +\left\{ b_{pq}^{-}e^{k_{pq}^{-}z}e^{i\theta_{pq}}+c.c.\right\} \nonumber \\
 &  & +\left\{ b_{rs}C_{rs}e^{k_{rs}z}e^{i\theta_{rs}}+c.c.\right\} .\label{eq: VP three waves}
\end{eqnarray}
Wherein, $b=(c-U)a$ and $\theta=kx-\omega t$. The first permutation
refers to absolute frequency and the second permutation denotes type
I or type II wavenumber modes (see $\mathsection$\ref{subsec: analytical disp}).
Linear evolution is given by (\ref{eq:linear operator-1}). 

\begin{eqnarray}
\mathscr{L}\left(\tilde{\phi}\right) & = & \frac{\partial^{2}\tilde{\phi}}{\partial t^{2}}+g\frac{\partial\tilde{\phi}}{\partial z}+2U\frac{\partial^{2}\tilde{\phi}}{\partial x\partial t}+U^{2}\frac{\partial^{2}\tilde{\phi}}{\partial x^{2}},\qquad z=0.\label{eq:linear operator}
\end{eqnarray}
The three-wave system in matrix form is

\begin{eqnarray}
\mathbf{M}b & = & 0.\label{eq: matrix form-1-1}
\end{eqnarray}
Wherein, $\mathbf{M}$ is a matrix of coefficients and $b$ is a vector
of amplitudes. Assuming (\ref{eq: VP three waves}), the coefficients
have the form

\begin{equation}
\mathscr{D}\left(k_{mn},\omega_{m},U\right)=U^{2}k_{mn}^{2}+\left(-2U\omega_{m}-g\right)k_{mn}+\omega_{m}^{2}=0.\label{eq: coefficients}
\end{equation}
Combining (\ref{eq: matrix form-1-1}) and (\ref{eq: coefficients})
yields
\begin{eqnarray}
\left[\begin{array}{ccc}
\mathscr{D}_{m}\left(k_{mn},\omega_{m},U\right) & 0 & 0\\
0 & \mathscr{D}_{p}\left(k_{pq}^{-},\omega_{p},-U\right) & 0\\
0 & 0 & \mathscr{D}_{r}\left(k_{rs},\omega_{r},U\right)
\end{array}\right]\left[\begin{array}{c}
b_{mn}\\
b_{pq}^{-}\\
b_{rs}
\end{array}\right] & = & \left[\begin{array}{c}
0\\
0\\
0
\end{array}\right].\label{eq: matrix form explicit-1-1}
\end{eqnarray}
A vanishing determinant yields the eigenvalues of the system

\begin{eqnarray}
\left(\left(\omega_{m}-k_{mn}U\right)^{2}-gk_{mn}\right)\left(\left(\omega_{p}+k_{pq}^{-}U\right)^{2}-gk_{pq}^{-}\right)\left(\left(\omega_{r}-k_{rs}U\right)^{2}-gk_{rs}\right) & = & 0.\label{eq: vanishing determinant 1}
\end{eqnarray}
This produces three dispersion relations

\begin{eqnarray}
\omega_{m}-k_{mn}U & = & \sqrt{gk_{mn}},\label{eq: wm1 disp}\\
\omega_{p}+k_{pq}^{-}U & = & \sqrt{gk_{pq}},\nonumber \\
\omega_{r}-k_{rs}U & = & \sqrt{gk_{rs}}.\label{eq: wr disp}
\end{eqnarray}
Seek a solution of (\ref{eq: wm1 disp})-(\ref{eq: wr disp}) which
also satisfies the constraints of a triad resonance,

\begin{eqnarray}
\omega_{m}+\omega_{p} & = & \omega_{r},\\
k_{mn}+k_{pq}^{-} & = & k_{rs}.
\end{eqnarray}
Recall, a positive wavenumber number convention has been chosen. Introduce
the scaling

\begin{alignat}{3}
\omega_{p} & \rightarrow & \mu\omega_{m},\qquad & \omega_{r} & \rightarrow & \left(1+\mu\right)\omega_{m}.\label{eq: scaling}
\end{alignat}
The resonance condition becomes

\begin{eqnarray}
\omega_{m}+\mu\omega_{m} & = & \left(1+\mu\right)\omega_{m},\label{eq: constraint scaled 1}\\
k_{mn}+k_{pq}^{-} & = & k_{rs}.\label{eq: k constraint 1}
\end{eqnarray}
$\mu$ is the ratio between initial harmonics. A self-interaction,
an interaction between two waves with same frequency, is given by
$\mu=1$. Increasing $\mu$ gives resonance for a higher upstream
harmonic. Combine (\ref{eq: vanishing determinant 1}) and (\ref{eq: scaling}),
and solve for wavenumbers. This yields
\begin{align}
k_{mn} & =\frac{g+2U\omega_{m}+\left(-1\right)^{n}\sqrt{g\left(g+4U\omega_{m}\right)}}{2U^{2}},\label{eq: kmn1 Dim}\\
k_{pq}^{-} & =\frac{g-2\mu U\omega_{m}+\left(-1\right)^{n}\sqrt{g(g-4\mu U\omega_{m})}}{2U^{2}},\label{eq: kpqminus}\\
k_{rs} & =\frac{g+2(\mu+1)U\omega_{m}+\left(-1\right)^{n}\sqrt{g\left(g+4(\mu+1)U\omega_{m}\right)}}{2U^{2}}.\label{eq: Krn dim}
\end{align}
If all modes are either type I or II modes $\left(n=q=s\right)$,
nontrivial solutions for (\ref{eq: kmn1 Dim})-(\ref{eq: Krn dim})
and (\ref{eq: k constraint 1}) exist. Figure \ref{fig:Schematic-of-possible}
gives a schematic of resonant combinations.

\begin{figure}
\centering{}\includegraphics[scale=0.35]{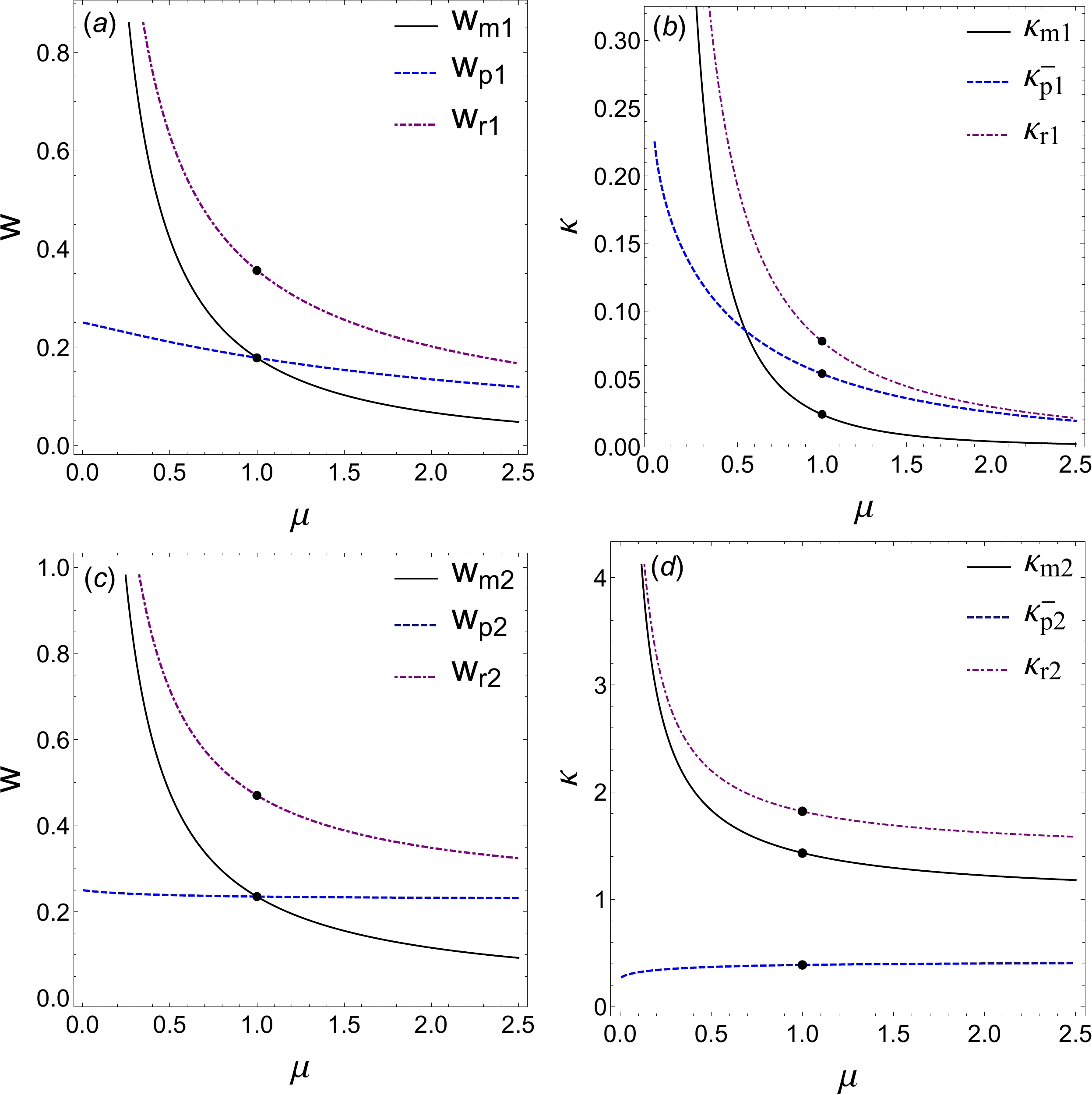}\caption{{\small{}Resonance solution curves for a laboratory viewer. The second
permutations, $n=1$ and $n=2$, correspond to type I and II wavenumber
modes. (a) Type I mode resonant frequencies: larger difference between
initial harmonics (larger $\mu$) generates slower waves (lower frequencies);
purple-dot dashed line. (b) Resonant type I wavenumbers: increasing
$\mu$ generates longer upstream waves (smaller wavenumbers). (c)
Type II mode resonant frequencies: larger difference between initial
harmonics (larger $\mu$) generates lower frequencies. (d) Resonant
type II wavenumbers: increasing $\mu$ generates longer waves. Black
dots mark resonant values for special case of a self-interaction $\left(\mu=1\right)$.
\label{fig: Resonance solutions curves - laboratory viewer}}}
\end{figure}

Resonance conditions reduce to a relationship between $\omega_{m}$
and $\mu$. Before satisfying the final constraint, (\ref{eq: k constraint 1}),
it is convenient to switch to non-dimensional forms,
\begin{alignat}{3}
\mathbf{\mathrm{w}} & = & \omega\frac{U}{g},\qquad & \kappa & = & k\frac{U^{2}}{g}.
\end{alignat}
The resonance closure becomes

\begin{eqnarray}
\mathrm{w}_{mn}+\mu\mathrm{w}_{mn}-\left(1+\mu\right)\mathrm{w}_{mn} & = & 0,\label{eq: w self-interaction closure}\\
\kappa_{mn}+\kappa_{pq}^{-}-\kappa_{rs} & = & 0.\label{eq: kappa self-interaction closure}
\end{eqnarray}
Wavenumber solutions are

\begin{flalign}
\kappa_{mn} & =\frac{1}{2}\left(1+2\mathrm{w}_{mn}+\left(-1\right)^{n}\sqrt{4\mathrm{w}_{mn}+1}\right),\label{eq: kappa1 lab viewer}\\
\kappa_{pq}^{-} & =\kappa_{pn}^{-}=\frac{1}{2}\left(1-2\mathrm{w}_{mn}\mu+\left(-1\right)^{n}\sqrt{1-4\mathrm{w}_{mn}\mu}\right),\label{eq: kappa 2 ab viewer}\\
\kappa_{rs} & =\kappa_{rn}=\frac{1}{2}\left(1+2\mathrm{w}_{mn}\left(1+\mu\right)+\left(-1\right)^{n}\sqrt{4\mathrm{w}_{mn}\left(1+\mu\right)+1}\right).\label{eq: kappa 3 lab viewer}
\end{flalign}
Equation (\ref{eq: kappa self-interaction closure}) and (\ref{eq: kappa1 lab viewer})-(\ref{eq: kappa 3 lab viewer})
reduce to

\begin{eqnarray}
4\mu\mathrm{w}_{mn}+\sqrt{1-4\mu\mathrm{w}_{mn}}+\sqrt{4\mathrm{w}_{mn}+1} & = & \sqrt{4(\mu+1)\mathrm{w}_{mn}+1}+1.\label{eq: gen closure}
\end{eqnarray}
Exact solution of (\ref{eq: gen closure}) is given in Appendix \ref{sec:Appendix exact solutions}.
If $\mu$ is known a priori, resonance conditions reduce to universal
constants. Some values are given in Table \ref{tab:Universal-constants-1}.
\begin{table}
\begin{centering}
\begin{tabular}{cccccc}
\toprule 
\multicolumn{6}{c}{\textbf{Type I modes $\mathbf{\left(n=1\right)}$}}\tabularnewline
\midrule
\midrule 
$\mu$ & .2 & .4 & .6 & .8 & 1\tabularnewline
\midrule 
$\mathrm{w}_{m1}=\frac{\omega U}{g}$ & 1.170 & .545 & .339 & .238 & .178\tabularnewline
\bottomrule
\end{tabular}$\qquad\qquad$%
\begin{tabular}{cccccc}
\toprule 
\multicolumn{6}{c}{\textbf{Type II modes $\mathbf{\left(n=2\right)}$}}\tabularnewline
\midrule
\midrule 
$\mu$ & .2 & .4 & .6 & .8 & 1\tabularnewline
\midrule 
$\mathrm{w}_{m2}=\frac{\omega U}{g}$ & 1.122 & .599 & .395 & .295 & .235\tabularnewline
\bottomrule
\end{tabular}
\par\end{centering}
\caption{{\small{}Deep water gravity wave triad resonance conditions reduce
to universal constants of non-dimensional parameter $\mathrm{w}_{mn}$.
Resonance closure is $\omega_{m}+\mu\omega_{m}=\left(1+\mu\right)\omega_{m}$,
$k_{mn}+k_{pq}^{-}=k_{rs}$. All modes must be type I or type II,
$n=q=s$. $\mu$ is ratio of initial harmonics. \label{tab:Universal-constants-1}}}

\end{table}
 Resonant frequency and wavenumber solution curves are shown in figure
\ref{fig: Resonance solutions curves - laboratory viewer}.
 For Type I modes, a greater difference in initial harmonics (larger
$\mu$) resonates lower frequencies and wavenumbers (see figs. \ref{fig: Resonance solutions curves - laboratory viewer}a-b).
Similar behavior is seen for type II mode resonant frequencies (see
fig. \ref{fig: Resonance solutions curves - laboratory viewer}c).
In contrast to the type I mode case, resonant upstream wavenumbers
are greater for type II modes (see fig. \ref{fig: Resonance solutions curves - laboratory viewer}d).

\subsection{Example: Self-interactions for absolute frequencies \label{subsec:Example-1-1}}

Consider a self-interaction $\left(\mu=1\right)$ between type I modes
$\left(n=1\right)$. The resonance condition, (\ref{eq: gen closure}),
is

\begin{equation}
\mathrm{w}_{m1}\approx.178.
\end{equation}
The exact form is given in Appendix \ref{sec:Appendix exact solutions}.
$\kappa$-values follow from (\ref{eq: kappa1 lab viewer})-(\ref{eq: kappa 3 lab viewer}).
Resonant frequencies and wavenumbers are

\begin{alignat}{2}
\mathrm{w}_{m1}\approx.178,\qquad & \mathrm{\mathrm{w}_{p1}}\approx.178,\qquad & \mathrm{w}_{r1}\approx.356,\label{eq: w closure type I ND}\\
\kappa_{m1}\approx.024,\qquad & \kappa_{p1}^{-}\approx.054,\qquad & \kappa_{r1}\approx.078.\label{eq: kappa closure type I ND}
\end{alignat}
Similarly, for type II modes $\left(n=2\right)$ and self-interactions
$\left(\mu=1\right)$,

\begin{equation}
\mathrm{w}_{m2}\approx.235.
\end{equation}
Resonant values are

\begin{alignat}{2}
\mathrm{w}_{m2}\approx.235,\qquad & \mathrm{\mathrm{w}_{p2}}\approx.235,\qquad & \mathrm{w}_{r2}\approx.470,\label{eq: w closure type II ND}\\
\kappa_{m2}\approx1.43,\qquad & \kappa_{p2}^{-}\approx.387,\qquad & \kappa_{r2}\approx1.820.\label{eq: kappa closure type II ND}
\end{alignat}
Numerical values may be found algebraically using algebraic procedures
or directly from the solution curves given in figure \ref{fig: Resonance solutions curves - laboratory viewer}.
Resonant values for the special case of self-interactions involving
type I and II modes, given in (\ref{eq: w closure type I ND})-(\ref{eq: kappa closure type I ND})
and (\ref{eq: w closure type II ND})-(\ref{eq: kappa closure type II ND})
respectively, are marked in Fig. \ref{fig: Resonance solutions curves - laboratory viewer}. 

Dimensional parameters are easily recovered from the non-dimensional
conditions. Some dimensional values are given in Table \ref{tab: Self-interactions-1}.
\begin{table}
\begin{centering}
\begin{tabular}{ccccccccccc}
\toprule 
 & \multicolumn{5}{c}{Type I} & \multicolumn{5}{c}{Type II}\tabularnewline
\midrule 
$\left|U\right|$ & $\omega_{m1}=\omega_{p1}^{-}$  & $\omega_{r1}$ & $k_{m1}$ & $k_{p1}^{-}$ & $k_{r1}$ & $\omega_{m2}=\omega_{p2}^{-}$  & $\omega_{r2}$ & $k_{m2}$ & $k_{p2}^{-}$ & $k_{r2}$\tabularnewline
\midrule
\midrule 
.5 & 3.49 & 6.98 & .94 & 2.12 & 3.10 & 4.61 & 9.21 & 56.06 & 15.29 & 71.34\tabularnewline
\midrule 
1 & 1.74 & 3.49 & .24 & .53 & .76 & 2.30 & 4.61 & 14.01 & 3.82 & 17.84\tabularnewline
\midrule 
1.5 & 1.16 & 2.33 & .10 & .24 & .34 & 1.54 & 3.07 & 6.23 & 1.70 & 7.93\tabularnewline
\midrule 
2 & .87 & 1.74 & .06 & .13 & .19 & 1.15 & 2.30 & 3.50 & .96 & 4.59\tabularnewline
\bottomrule
\end{tabular}
\par\end{centering}
\caption{{\small{}Dimensional values which satisfy self-interaction resonance
conditions as observed by a laboratory viewer. Self-interactions satisfy
universal values of non-dimensional parameters, $\left\{ \kappa_{m1},\kappa_{p1}^{-},\kappa_{r1},\mathrm{w}_{m}\right\} \approx\left\{ .024,.054,.078,.178\right\} $
and $\left\{ \kappa_{m2},\kappa_{p2}^{-},\kappa_{r2},\mathrm{w}_{m}\right\} \approx\left\{ 1.43,.387,1.82,.235\right\} $
for type I and II modes respectively. Generally speaking, faster currents
resonate lower frequencies (longer waves). The gravity wave regime
is most affected. (Units: $U\left[\frac{m}{s}\right]$, $k\left[\frac{rad}{m}\right]$
and $\omega\left[\frac{rad}{s}\right]$ ) \label{tab: Self-interactions-1}}}
\end{table}
 Realistic current velocities mostly affect the gravity wave regime.
Generally, faster currents resonate lower frequencies (longer waves)
and slower currents resonate higher frequencies (shorter waves). The
existence of two different resonant conditions suggests the dominant
energy transfer mechanism will not be homogeneous across a wave spectrum.
The non-dimensional parameter $\mathrm{w}=\omega Ug^{-1}$ has been
discussed for resonance between waves and an external forcing \citep[cf.][]{dagan1982free,tyvand2012surface}.
The above results suggest the parameter is also an indicator of an
intrinsic energy sharing mechanism within wave-current fields. 

\section{Resonance for an observer moving with velocity U\label{sec: 5 Res in comoving}}

In $\mathsection$\ref{sec: 4 Res in lab}, resonance conditions were
derived for a laboratory observer. It is temping to assume, since
applying a Galilean shift to a single plane wave yields the intrinsic
frequency of that wave, uniform application of a Galilean shift to
the resonance closure in a lab frame will recover the resonance closure
in terms of intrinsic frequencies. This is not always true. Indeed,
for the triad resonance between waves propagating in opposite directions
with respect to current, discussed in section $\mathsection$\ref{sec: 4 Res in lab},
it is false (see Appendix \ref{sec: matrix arguments}). A rigorous
transformation to a moving observer ought to derive the dispersion
relations from the governing equations anew.

Define a second observer, moving with a velocity $U'$, who observes
the resonant system discussed in $\mathsection$\ref{sec: 4 Res in lab}.
One may not assume $a$ $priori$ that this observer describes the
system by three intrinsic frequencies (see Appendix \ref{sec: matrix arguments}).
Denote frequencies for this viewer by a yet to be defined frequency
description: $\upsilon$-frequencies. The resonance closure in terms
of $\upsilon$-frequencies is

\begin{eqnarray}
\upsilon_{m}+\upsilon_{p}-\upsilon_{r} & = & 0,\label{eq: frequency closure (intrinsic - dimensional)-1}\\
k_{m}+k_{p}^{-}-k_{r} & = & 0.\label{eq: wavenumber closure (intrinsic - dimensional)-1}
\end{eqnarray}
It should be emphasized, the relationship between \ref{eq: frequency closure (intrinsic - dimensional)-1}
and \ref{eq: wavenumber closure} is given by the dispersion relations
which are not known $a$ $priori$. The procedure of $\mathsection$\ref{sec: 4 Res in lab}
may now be repeated. Dispersion relations are derived assuming periodicity,
and resonance conditions are found by solving the system's determinant
along with a constraint for resonant phase matching.

The linear operator, (\ref{eq:linear operator-1}), used in $\mathsection$\ref{sec: 4 Res in lab},
may be applied for the new observer. The evolution equation for each
plane wave is now given by

\begin{eqnarray}
\mathscr{L}\left(\tilde{\phi}\right) & = & \frac{\partial^{2}\tilde{\phi}}{\partial t^{2}}+g\frac{\partial\tilde{\phi}}{\partial z}+2V_{mn}\frac{\partial^{2}\tilde{\phi}}{\partial x\partial t}+V_{mn}^{2}\frac{\partial^{2}\tilde{\phi}}{\partial x^{2}},\qquad z=0.\label{eq:linear operator-2}
\end{eqnarray}
Wherein, $V_{mn}$ is an effective current on each plane wave in the
new description. It is found by addition of velocities,

\begin{eqnarray*}
V_{mn} & = & \pm U-U'.
\end{eqnarray*}
If the velocity of the moving observer is the same sign and magnitude
as current, $U'=U$. Then, for waves following current, $V_{mn}=+U-U=0$.
And, for the wave opposing the current, $V_{mn}^{-}=-U-U=-2U$. Thus,
the two waves following current (and observer), $a_{rs}$ and $a_{mn}$,
will ``feel'' no effective current. Whereas, the wave opposing current
(and observer), $a_{pq}^{-}$, will ``feel'' an ambient current
due to the preferential direction of the media. Using these definitions
for $V_{mn}$, one may now derive the correct dispersion relations
from the governing equations. A less conceptual, but more mathematically
transparent, derivation is given in Appendix \ref{sec: matrix arguments}.

The three-wave system in matrix form is

\begin{eqnarray}
\mathbf{M}b & = & 0.\label{eq: matrix form-1}
\end{eqnarray}
Wherein, $\mathbf{M}$ is a matrix of coefficients and $b$ is a vector
of velocity potential amplitudes. Assume spatial and temporal periodicity,
the coefficients become

\begin{equation}
\mathscr{D}\left(k_{mn},\upsilon_{m},V_{mn}\right)=V_{mn}^{2}k_{mn}^{2}+\left(-2V_{mn}\upsilon_{m}-g\right)k_{mn}+\upsilon_{m}^{2}=0.\label{eq: wc dispersion-2-1-1}
\end{equation}
Combine (\ref{eq: matrix form-1}) and (\ref{eq: wc dispersion-2-1-1})
with appropriate definitions of $V$,
\begin{eqnarray}
\left[\begin{array}{ccc}
\mathscr{D}_{m}\left(k_{mn},\upsilon_{m},0\right) & 0 & 0\\
0 & \mathscr{D}_{p}\left(k_{pq}^{-},\upsilon_{p},-2U\right) & 0\\
0 & 0 & \mathscr{D}_{r}\left(k_{rs},\upsilon_{m},0\right)
\end{array}\right]\left[\begin{array}{c}
b_{mn}\\
b_{pq}\\
b_{rs}
\end{array}\right] & = & \left[\begin{array}{c}
0\\
0\\
0
\end{array}\right].\label{eq: matrix form explicit-1}
\end{eqnarray}
Eigenvalues are found when the determinant of (\ref{eq: matrix form explicit-1})
vanishes,

\begin{eqnarray}
\left(\ensuremath{\upsilon_{m}^{2}}-gk_{mn}\right)\left(\left(\ensuremath{\upsilon_{p}+2k_{pq}^{-}U}\right)^{2}-gk_{pq}\right)\left(\ensuremath{\upsilon_{r}^{2}}-gk_{rs}\right) & = & 0.\label{eq: Determinant-1}
\end{eqnarray}
This produces three dispersion relations

\begin{eqnarray}
\upsilon_{m} & = & \sqrt{gk_{m}},\label{eq: ups_m disp}\\
\upsilon_{p}+2k_{p}^{-}U & = & \sqrt{gk_{p}^{-}},\\
\upsilon_{r}^{2} & = & \sqrt{gk_{r}}.\label{eq: ups_r disp}
\end{eqnarray}
Seek a solution of (\ref{eq: ups_m disp})-(\ref{eq: ups_r disp})
which also satisfies the constraint of resonance,
\begin{eqnarray}
\upsilon_{m}+\mu\upsilon_{m}-\left(1+\mu\right)\upsilon_{m} & = & 0,\label{eq: upsilon closure-1}\\
k_{mn}+k_{pq}^{-}-k_{rs} & = & 0.\label{eq: 511}
\end{eqnarray}
Combine (\ref{eq: Determinant-1}) and (\ref{eq: upsilon closure-1}),
taking into account the resonance requirement that $n=q=s$, yields
the wavenumbers,

\begin{flalign}
k_{mn} & =\frac{\upsilon_{m}^{2}}{g},\label{eq: km1 v D}\\
k_{pq}^{-} & =k_{pn}^{-}=\frac{g-4\mu U\upsilon_{m}+\left(-1\right)^{n}\sqrt{g(g-8\mu U\upsilon_{m})}}{8U^{2}},\label{eq: kp1 v D}\\
k_{rs} & =k_{rn}=\frac{(\mu+1)^{2}\upsilon_{m}^{2}}{g}.\label{eq: kr1 v D}
\end{flalign}
The dispersion relations for the two waves propagating with the current,
(\ref{eq: km1 v D}) and (\ref{eq: kr1 v D}), are single valued functions.
Whereas, the dispersion relation for the plane wave propagating against
the current, (\ref{eq: kp1 v D}), is multi-valued. It is of interest
to note, the third term in (\ref{eq: kp1 v D}) gives the wave-blocking
condition, $g-8\mu U\upsilon_{m}=0$,

\begin{eqnarray}
\frac{U\upsilon_{m}}{g} & > & \frac{1}{8\mu}.
\end{eqnarray}
Noting $U=\nicefrac{V}{2}$ recovers the known blocking condition
$\sim\nicefrac{1}{4}$. Nontrivial solutions to (\ref{eq: 511}) and
(\ref{eq: km1 v D})-(\ref{eq: kr1 v D}) are found when all modes
are either type I or type II $\left(n=q=s\right)$. Note, the second
permutation is retained for all waves even when their dispersion form
has one solution. Non-dimensionalization shows all solutions are implicitly
dependent on the permutation $n$. See figure \ref{fig:Schematic-of-possible}b
for a schematic of resonant wavenumber combinations. Before satisfying
(\ref{eq: 511}), introduce non-dimensional parameters

\begin{alignat}{3}
\mathrm{v} & = & \upsilon\frac{U}{g},\qquad & \kappa & = & k\frac{U^{2}}{g}.
\end{alignat}
Wavenumber solutions become

\begin{flalign}
\kappa_{mn} & =\mathrm{v}_{m}^{2},\label{eq: km1 v ND}\\
\kappa_{pq}^{-} & =k_{pn}^{-}=\frac{1}{8}\left(1-4\mu\mathrm{v}_{m}+\left(-1\right)^{n}\sqrt{1-8\mu\mathrm{v}_{m}}\right),\label{eq: kp1 v ND}\\
\kappa_{rs} & =\kappa_{rn}=(\mu+1)^{2}\mathrm{v}_{m}^{2}.\label{eq: kr1 v ND}
\end{flalign}
Equations (\ref{eq: km1 v ND})-(\ref{eq: kr1 v ND}) and (\ref{eq: 511})
reduce to

\begin{eqnarray}
\frac{1}{8}\left(1+(-1)^{n}\sqrt{1-8\mu\mathrm{v}_{mn}}-4\mu\mathrm{v}_{mn}(1+2(\mu+2)\mathrm{v}_{mn})\right) & = & 0.\label{eq: comoving res cond}
\end{eqnarray}
Equation (\ref{eq: comoving res cond}) has an exact solution when

\begin{eqnarray}
\mathrm{v}_{mn} & = & \left(\mu(\mu+2)+\left(-1\right)^{n+1}\sqrt{\mu(\mu+2)^{3}}\right)^{-1}.\label{eq: v exact sol}
\end{eqnarray}
Substitute (\ref{eq: v exact sol}) into (\ref{eq: km1 v D})-(\ref{eq: kr1 v D}).
Resonant wavenumbers are
\begin{eqnarray}
\kappa_{mn} & = & \frac{1}{\left(\mu(\mu+2)+\left(-1\right)^{n+1}\sqrt{\mu(\mu+2)^{3}}\right)^{2}},\label{eq: kmn res}\\
\kappa_{pn}^{-} & = & \frac{1}{8}\left(1-\frac{4\mu}{\mu(\mu+2)+\left(-1\right)^{n+1}\sqrt{\mu(\mu+2)^{3}}}\right.\nonumber \\
 &  & \left.\qquad+\left(-1\right)^{n}\sqrt{1-\frac{8\mu}{\mu(\mu+2)+\left(-1\right)^{n+1}\sqrt{\mu(\mu+2)^{3}}}}\right),\label{eq:kpn res}\\
\kappa_{rn} & = & \frac{(\mu+1)^{2}}{\left(\mu(\mu+2)+\left(-1\right)^{n+1}\sqrt{\mu(\mu+2)^{3}}\right)^{2}}.\label{eq: krn res}
\end{eqnarray}
When $n=1$, (\ref{eq:kpn res}) gives the wave-blocking condition
in $\mu$-space. The blocking point occurs when 

\[
1-\frac{8\mu}{\mu(\mu+2)+\sqrt{\mu(\mu+2)^{3}}}=0.
\]
This corresponds to 

\begin{eqnarray}
\mu & = & \frac{2}{3}.\label{eq: blocking point}
\end{eqnarray}

\begin{figure}
\begin{centering}
\includegraphics[scale=0.35]{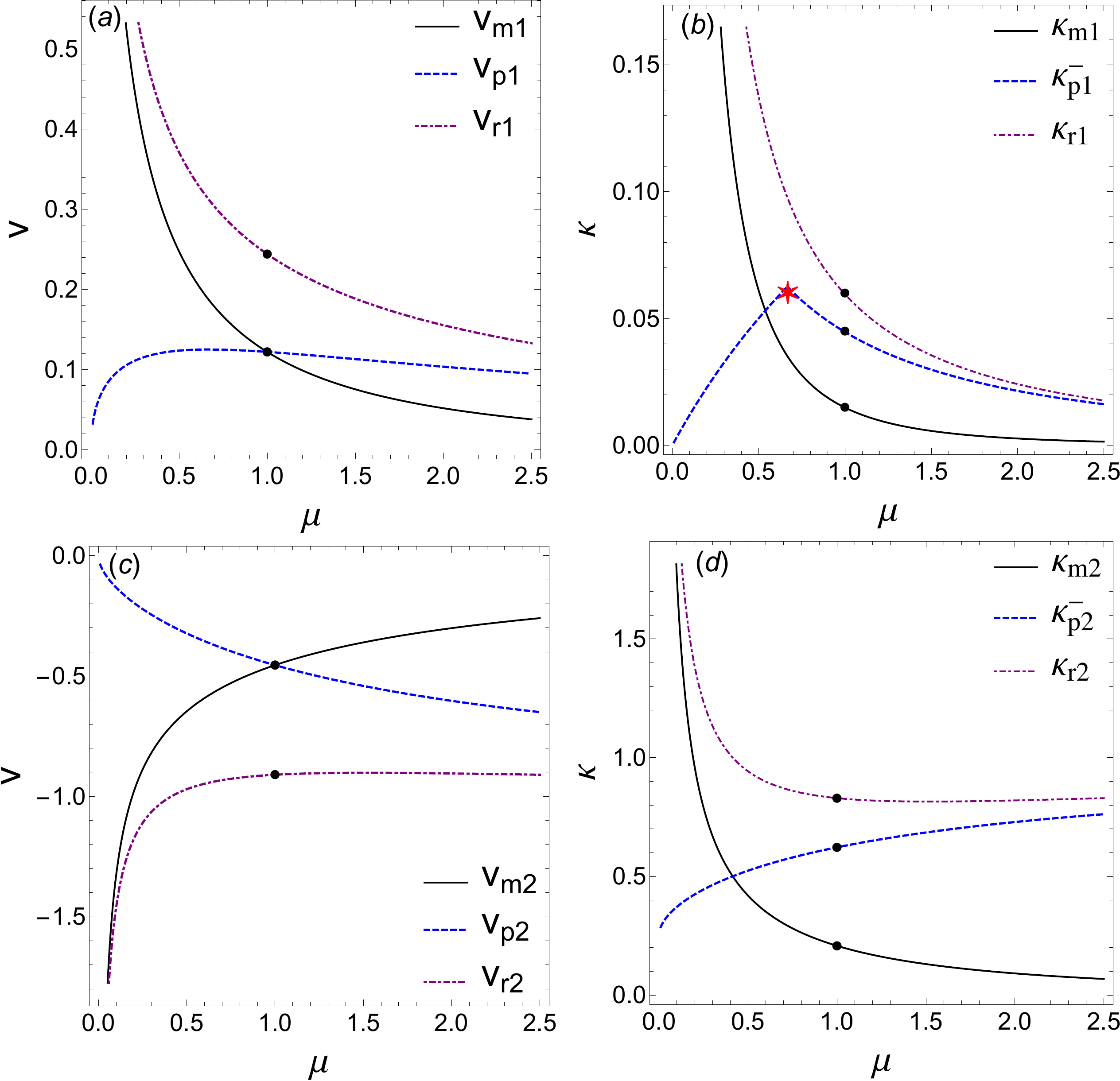}
\par\end{centering}
\centering{}\caption{{\small{}Resonant solution curves for antiparallel waves seen by an
observer moving with the current velocity $U$. The second permutations,
$n=1$ and $n=2$, correspond to type I and II wavenumber modes. (a)
Type I mode resonant frequencies: larger difference in initial harmonics
(larger $\mu$) generates lower frequencies, purple-dot dashed line.
(b) Resonant type I wavenumbers: increasing $\mu$ generates long
waves (lower wavenumbers). Effect of wave-blocking on resonances seen
at inflection point; red star, of upstream wave solution curve; blue-dashed
line. (c) Type II mode resonant frequencies: larger $\mu$ generates
smaller negative frequencies. (d) Resonant type II wavenumbers: increasing
$\mu$ generates longer waves. Black dots mark resonant values for
special case of a self-interaction $\left(\mu=1\right)$. \label{fig:Resonance solution curves - comoving observer}}}
\end{figure}

The $v$-description is more analytically tractable than the laboratory
frame (compare (\ref{eq: v exact sol}) with Appendix \ref{sec:Appendix exact solutions}).
Thus, it is more convenient for analysis. Solution curves for resonant
values measured by the moving observer are given in figure \ref{fig:Resonance solution curves - comoving observer}.
 Resonant type I frequencies are positive and decrease with $\mu$
(see fig. \ref{fig:Resonance solution curves - comoving observer}a).
The solution curve for resonant type I wavenumbers exhibits an inflection
point at the wave-blocking condition $\mu=\nicefrac{2}{3}$ (see fig.
\ref{fig:Resonance solution curves - comoving observer}b). The inflection
point occurs for the upstream wave. For type II modes, the resonant
frequencies are negative (see fig. \ref{fig:Resonance solution curves - comoving observer}c).
A greater difference between initial harmonics (larger $\mu$) generates
longer waves (see fig. \ref{fig:Resonance solution curves - comoving observer}d).

\subsection{Example: Self-interactions\label{subsec:Example-1}}

For a self-interaction, $\mu=1$ in (\ref{eq: v exact sol}) and

\begin{eqnarray}
\mathrm{v}_{mn} & = & \left(3-\left(-1\right)^{n}3\sqrt{3}\right)^{-1}.\label{eq: v n=00003D1}
\end{eqnarray}
Consider a self-interaction between type I modes $\left(n=1\right)$,
(\ref{eq: v n=00003D1}) yields

\begin{flalign}
\mathrm{v}_{n1} & =\left(3+3\sqrt{3}\right)^{-1}\approx.122.
\end{flalign}
Equations (\ref{eq: kmn res})-(\ref{eq: krn res}) give resonant
$\kappa$-values. Resonance values are

\begin{alignat}{2}
\mathrm{v}_{m1}\approx.122,\qquad & \mathrm{\mathrm{v}_{p1}}\approx.122,\qquad & \mathrm{v}_{r1}\approx.244,\label{eq: v closure type I ND}\\
\kappa_{m1}\approx.015\qquad & \kappa_{p1}^{-}\approx.045,\qquad & \kappa_{r1}\approx.060.\label{eq:kappa closure type I ND}
\end{alignat}
Similarly, for type II modes $\left(n=2\right)$ and self-interactions
$\left(\mu=1\right)$,

\begin{flalign}
\mathrm{v}_{n2} & =\left(3-3\sqrt{3}\right)^{-1}\approx-.455.
\end{flalign}
Resonant values are

\begin{alignat}{2}
\mathrm{v}_{m2}\approx-.455,\qquad & \mathrm{\mathrm{v}_{p2}}\approx-.455,\qquad & \mathrm{v}_{r2}\approx-.910,\label{eq: v closure type II ND}\\
\kappa_{m2}\approx.207\qquad & \kappa_{p2}^{-}\approx.622,\qquad & \kappa_{r2}\approx.829.\label{eq: kappa closure type II ND-1}
\end{alignat}
Numerical values may be found algebraically or directly from the solution
curves given in figure \ref{fig:Resonance solution curves - comoving observer}.
Resonant values for the special case of self-interactions involving
type I and II modes, given in (\ref{eq: v closure type I ND})-(\ref{eq:kappa closure type I ND})
and (\ref{eq: v closure type II ND})-(\ref{eq: kappa closure type II ND-1})
respectively, are marked by black dots in Fig. \ref{fig:Resonance solution curves - comoving observer}.

\section{Amplitude coupling \label{sec: Amplitude coupling and energy exchange}}

Growth due to resonant phase matching is known to be arrested by amplitude
coupling \citep{benney1962non}. Nonetheless, if the interaction time
is sufficiently long, nonlinear terms may still grow to first-order
magnitude. In this section, a three-wave interaction model for surface
amplitudes is adapted for a resonant wave-current system involving
waves propagating in both directions with respect to current.

It was discussed in $\mathsection$\ref{sec: 5 Res in comoving} that
a viewer moving with a velocity $U$, who will see the fluid ``still,''
will not describe the system of waves propagating in opposite directions
with respect to current, given in $\mathsection$\ref{sec: 4 Res in lab},
by three intrinsic frequencies (see also Appendix \ref{eq: apply gal shift}).
Rather, this observer sees frequencies given by the dispersion relations
(\ref{eq: ups_m disp})-(\ref{eq: ups_r disp}). Incorporating the
definitions of the intrinsic frequency, $\sigma_{i}=\sqrt{gk_{i}}$
into these dispersion relations yields

\begin{eqnarray}
\upsilon_{m} & = & \sigma_{m},\label{eq: nu to sigam trans 1}\\
\upsilon_{p} & = & \sigma_{p}-2k_{p}^{-}U,\\
\upsilon_{r} & = & \sigma_{r}.\label{eq: nu to sigma trans 3}
\end{eqnarray}
Thus, $\upsilon$-frequencies may be understood to be the intrinsic
frequencies of the original system transformed in such a way that
the mode propagating against current has a correction due to the preferential
direction of the media. As such, it is the $\upsilon$-description,
not the $\sigma$-description, which consistently describe a system
of waves propagating on both directions with respect to current for
a viewer who moves with the velo. Accordingly, the transformations
given by (\ref{eq: nu to sigam trans 1})-(\ref{eq: nu to sigma trans 3}),
may be incorporated into a three-wave interaction model for a viewer
which sees still water.

A coupled three-wave model for a viewer who sees surface water waves
propagating on ``still'' water is of the form

\begin{equation}
\frac{da_{r}}{d\tau}=\beta^{\left(1\right)}a_{m}a_{p},\qquad\frac{da_{m}}{d\tau}=-\beta^{\left(2\right)}a_{p}a_{r},\qquad\frac{da_{p}}{d\tau}=-\beta^{\left(3\right)}a_{m}a_{r}.\label{eq: three-wave system-1}
\end{equation}
Wherein, the nonlinear coefficients are

\begin{eqnarray}
\beta^{\left(1\right)} & = & \frac{k_{r}\upsilon_{m}\upsilon_{p}}{4\upsilon_{r}^{2}}\left(\frac{k_{m}\upsilon_{r}^{2}}{k_{r}\upsilon_{p}}+\frac{k_{p}\upsilon_{r}^{2}}{k_{r}\upsilon_{m}}-\frac{\upsilon_{m}^{2}}{\upsilon_{p}}-\frac{\upsilon_{p}^{2}}{\upsilon_{m}}\right),\label{eq: Alpha (B) 12}\\
\beta^{\left(2\right)} & = & \frac{k_{m}\upsilon_{p}\upsilon_{r}}{4\upsilon_{m}^{2}}\left(\frac{k_{r}\upsilon_{m}^{2}}{k_{m}\upsilon_{p}}-\frac{k_{p}\upsilon_{m}^{2}}{k_{m}\upsilon_{r}}-\frac{\upsilon_{r}^{2}}{\upsilon_{p}}+\frac{\upsilon_{p}^{2}}{\upsilon_{r}}+4\upsilon_{r}\right),\label{eq: Alpha (B) 23}\\
\beta^{\left(3\right)} & = & \frac{k_{p}\upsilon_{m}\upsilon_{r}}{4\upsilon_{p}^{2}}\left(-\frac{k_{m}\upsilon_{p}^{2}}{k_{p}\upsilon_{r}}+\frac{k_{r}\upsilon_{p}^{2}}{k_{p}\upsilon_{m}}+\frac{\upsilon_{m}^{2}}{\upsilon_{r}}-\frac{\upsilon_{r}^{2}}{\upsilon_{m}}+4\upsilon_{p}\right).\label{eq: Alpha (B) 13}
\end{eqnarray}
The reader is referred to \citet{mcgoldrick1965resonant} for the
original derivation. The above form is reduced for the case of  waves
without capillary effects. In contrast to the original formulation,
it is written in terms of $\upsilon$-frequencies which are $\sigma$-frequencies
with a correction for wave motion in both directions with respect
to current.

Introduce non-dimensional parameters

\begin{alignat}{2}
\mathrm{v}_{mn}=\frac{\upsilon_{m}U}{g},\qquad & \kappa_{mn}=\frac{k_{mn}U^{2}}{g},\qquad & \tau=\upsilon_{ref}t,\qquad & A_{mn}=k_{ref}a_{mn}.\label{eq: ND params for model}
\end{alignat}
$\upsilon_{ref}$ and $k_{ref}$ are the frequency and wavenumber
of the initial component following current. The system may be written
as

\begin{equation}
\frac{d\mathrm{A}_{rn}}{d\tau}=\mathrm{B}^{\left(mp\right)}\mathrm{A}_{mn}\mathrm{A}_{pq}^{-},\qquad\frac{d\mathrm{A}_{mn}}{d\tau}=-\mathrm{B}^{\left(pr\right)}\mathrm{A}_{pq}^{-}\mathrm{A}_{rs},\qquad\frac{d\mathrm{A}_{pq}^{-}}{d\tau}=-\mathrm{B}^{\left(mr\right)}\mathrm{A}_{mn}\mathrm{A}_{rs}.\label{eq: three-wave system}
\end{equation}
Nonlinear coefficients are

\begin{alignat}{1}
\mathrm{B}^{\left(mp\right)} & =\frac{\kappa_{rs}\mathrm{v}_{mn}\mathrm{v}_{pq}}{4\mathrm{v}_{rs}^{2}}\left(\frac{\mathrm{v}_{rs}^{2}}{\kappa_{rs}\mathrm{v}_{mn}\mathrm{v}_{pq}}+\frac{\kappa_{pq}\mathrm{v}_{rs}^{2}}{\kappa_{rs}\kappa_{mn}\mathrm{v}_{mn}^{2}}-\frac{\mathrm{v}_{mn}}{\kappa_{mn}\mathrm{v}_{pq}}-\frac{\mathrm{v}_{pq}^{2}}{\kappa_{mn}\mathrm{v}_{mn}^{2}}\right),\label{eq: Coeff1 ND}\\
\mathrm{B}^{\left(pr\right)} & =\frac{\kappa_{mn}\mathrm{v}_{rs}\mathrm{v}_{pq}}{4\mathrm{v}_{mn}^{2}}\left(\frac{\mathrm{v}_{mn}^{2}\kappa_{rs}}{\mathrm{v}_{pq}^{2}\kappa_{mn}\kappa_{pq}}-\frac{\mathrm{v}_{mn}^{2}}{\mathrm{v}_{rs}\kappa_{mn}\mathrm{v}_{pq}}-\frac{\mathrm{v}_{rs}^{2}}{\mathrm{v}_{pq}^{2}\kappa_{pq}}+\frac{\mathrm{v}_{pq}}{\mathrm{v}_{rs}\kappa_{pq}}+\frac{4\mathrm{v}_{rs}}{\mathrm{v}_{pq}\kappa_{pq}}\right),\label{eq: Coeff2 ND}\\
\mathrm{B}^{\left(mr\right)} & =\frac{\kappa_{pq}\mathrm{v}_{mn}\mathrm{v}_{rs}}{4\mathrm{v}_{pq}^{2}}\left(-\frac{\kappa_{mn}}{\kappa_{pq}\kappa_{rs}\mathrm{v}_{rs}^{2}}+\frac{\mathrm{v}_{pq}^{2}}{\kappa_{pq}\mathrm{v}_{mn}\mathrm{v}_{rs}}+\frac{\mathrm{v}_{mn}^{2}}{\kappa_{rs}\mathrm{v}_{rs}^{2}}-\frac{\mathrm{v}_{rs}}{\kappa_{rs}\mathrm{v}_{mn}}+\frac{4\mathrm{v}_{pq}^{2}}{\mathrm{v}_{pq}\kappa_{rs}\mathrm{v}_{rs}}\right).\label{eq: Coeff3 ND}
\end{alignat}
For resonant waves ($n=q=s$), wavenumbers are given by (\ref{eq: kmn res})-(\ref{eq: krn res}).
Equations (\ref{eq: Coeff1 ND})-(\ref{eq: Coeff3 ND}) become

\begin{eqnarray}
\mathrm{B}_{n}^{\left(mp\right)} & = & \frac{1}{32\mathrm{v}_{mn}^{2}}\left(\left(1+\left(-1\right)^{n}R_{mn}\right)\mu-4\mathrm{\mathrm{v}_{mn}}\mu^{2}-8\mathrm{v}_{mn}^{2}\mu^{3}\right),\label{eq: Coeff1}\\
\mathrm{B}_{n}^{\left(pr\right)} & = & -\frac{1}{4}+\frac{2\mathrm{v_{mn}^{2}}\left(4+8\mu+5\mu^{2}\right)}{1+\left(-1\right)^{n}R_{mn}-4\mathrm{v_{mn}\mu}},\label{eq: Coeff2}\\
\mathrm{B}_{n}^{\left(mr\right)} & = & \frac{1}{32\mu^{2}(\mu+1)\mathrm{v}_{mn}^{2}}\left(1+\left(-1\right)^{n}+\left(1+(-1)^{n}R_{mn}-4v\right)\mu+\left(-(-1)^{n}R_{mn}\right.\right.\nonumber \\
 &  & \left.\left.+24\mathrm{v}_{mn}^{2}-4\mathrm{v}_{mn}-1\right)\mu^{2}+\left(24\mathrm{v}_{mn}^{2}+4\mathrm{v}_{mn}\right)\mu^{3}+8\mathrm{v}_{mn}^{2}\mu^{4}\right).\label{eq: Coeff3}
\end{eqnarray}
Wherein,

\begin{eqnarray*}
R_{mn} & = & \sqrt{1-8\mu\mathrm{v}_{mn}}.
\end{eqnarray*}
Equations (\ref{eq: Coeff1})-(\ref{eq: Coeff3}) depend only on $\mu$,
$\mathrm{v}$ and permutation $n$. However, $\mathrm{v}$ is also
expressible as a function solely of $\mathrm{\mu}$ through (\ref{eq: v exact sol}).
Accordingly, nonlinear coefficients have been written solely as a
function of $\mu$ and the permutation $n$, i.e. $\mathrm{B}_{n}^{\left(mp\right)}=\mathrm{B}_{n}^{\left(mp\right)}\left(\mu\right)$.

\begin{figure}
\centering{}\includegraphics[scale=0.35]{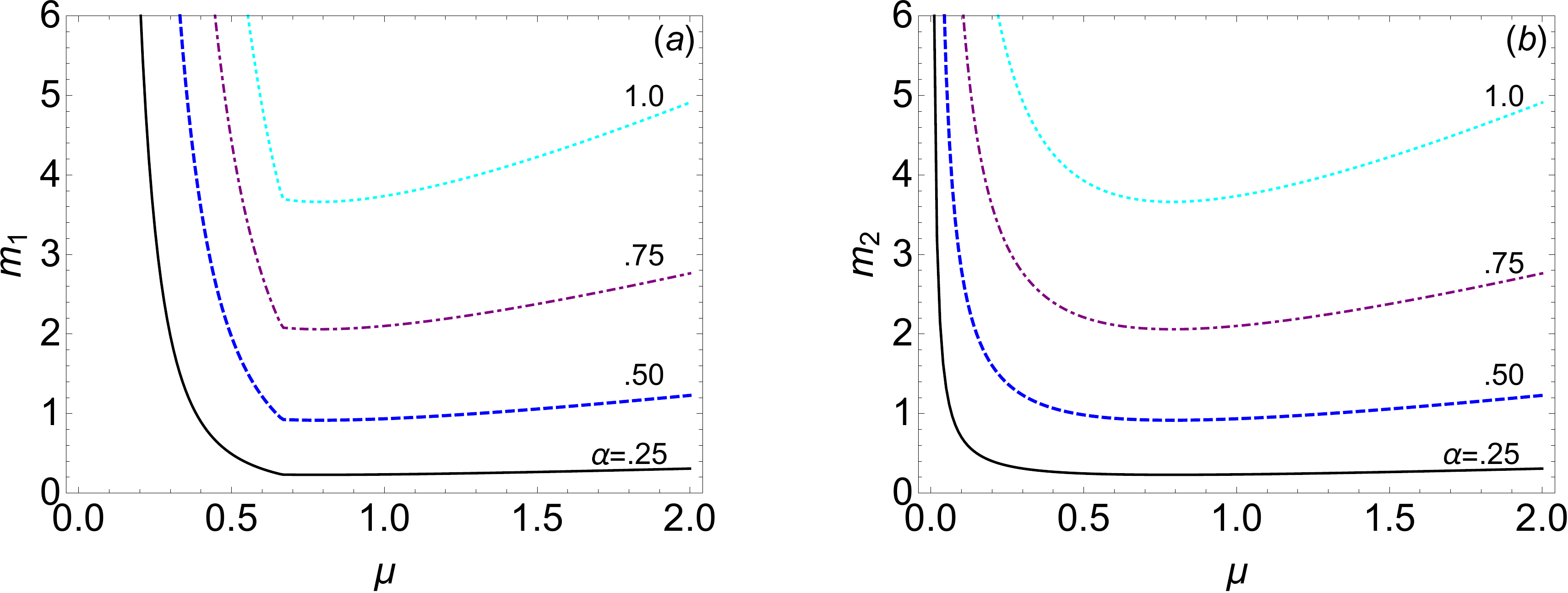}\caption{{\small{}Effect of $\mu$ on parameter $\mathrm{m}.$ Various initial
conditions of $\alpha=\hat{\mathrm{A}}_{p1}/\hat{\mathrm{A}}_{m1}$
shown for (a) type I modes and (b) type II modes. When $\mathrm{m}\protect\leq1$,
the $\kappa_{pq}^{-}$ mode transfers energy to the generated wave
at a faster rate than the $\kappa_{mn}$ mode. When $\mathrm{m}>1$,
the roles are reversed. The $\kappa_{mn}$ mode transfers energy to
the generated wave at a faster rate than the $\kappa_{pq}$ mode.
For $\mu<\nicefrac{2}{3}$, $\mathrm{m}\protect\leq1$ for type I
modes but $\mathrm{m}>1$ for type II modes. In this regime, energy
will be transferred more effectively from different directions on
different wavenumber scales. The upper bound of this regime, $\mu=\nicefrac{2}{3}$,
corresponds to the blocking point (see (\ref{eq: blocking point})).
\label{fig: m dependency}}}
\end{figure}

A solution to the system is known in terms of Jacobi-Elliptic functions
(see \citealp{mcgoldrick1965resonant,alam2010oblique}). Define initial
amplitudes: $\mathrm{A}_{mn}\left(0\right)=\hat{\mathrm{A}}_{mn}\neq0$,
$\mathrm{A}_{pn}^{-}\left(0\right)=\hat{\mathrm{A}}_{pn}^{-}\neq0$
and $\mathrm{A}_{rn}\left(0\right)=0$. Let $\alpha$ be a ratio of
the initial upstream and downstream amplitudes, $\alpha_{n}=\nicefrac{\hat{\mathrm{A}}_{pn}^{-}}{\hat{\mathrm{A}}_{mn}}$.
The solution is governed by the parameter

\begin{eqnarray}
\mathrm{m}_{n} & = & \frac{\mathrm{B}_{n}^{\left(pr\right)}\left(\hat{\mathrm{A}}_{pn}^{-}\right)^{2}}{\mathrm{B}_{n}^{\left(mr\right)}\hat{\mathrm{A}}_{mn}^{2}}.\label{eq: m parameter}
\end{eqnarray}
The dependency of $\mathrm{m}_{n}$ on $\mu$ and initial amplitudes
is shown in figure \ref{fig: m dependency}.

There are two solution cases. First, $\mathrm{m}_{n}\leq1$, the resonant
amplitude is

\begin{eqnarray}
\mathrm{A}_{rn}\left(\tau\right) & = & \hat{\mathrm{A}}_{pn}^{-}\left(\frac{\mathrm{B}_{n}^{\left(mp\right)}}{\mathrm{B}_{n}^{\left(mr\right)}}\right)^{\nicefrac{1}{2}}\mathrm{sn}\left(\mathrm{\chi_{n}}|\mathrm{m}_{n}\right).\label{eq: Ars  m<=00003D1}
\end{eqnarray}
Wherein, $\chi_{n}=\hat{\mathrm{A}}_{mn}^{2}\left(\mathrm{B}_{n}^{\left(mp\right)}\mathrm{B}_{n}^{\left(mr\right)}\right)^{\nicefrac{1}{2}}\left(\tau-\tau_{0}\right)$.
The initial waves evolve as

\begin{eqnarray}
\mathrm{A}_{mn}\left(\tau\right) & = & \hat{\mathrm{A}}_{mn}\mathrm{dn}\left(\chi_{n}|\mathrm{m}_{n}\right),\\
\mathrm{A}_{pn}^{-}\left(\tau\right) & = & \hat{\mathrm{A}}_{pn}^{-}\mathrm{cn}\left(\chi_{n}|\mathrm{m}_{n}\right).
\end{eqnarray}
The second case is $\mathrm{m}_{n}>1$, the resonant amplitude is

\begin{eqnarray}
\mathrm{A}_{rn}\left(\tau\right) & = & \hat{\mathrm{A}}_{mn}\left(\frac{\mathrm{B}_{n}^{\left(mp\right)}}{\mathrm{B}_{n}^{\left(pr\right)}}\right)^{\nicefrac{1}{2}}\mathrm{sn}\left(\chi_{n}|\mathrm{m}_{n}^{-1}\right).\label{eq: a3s m<=00003D1-1}
\end{eqnarray}
Wherein, $\chi_{n}=\left(\hat{\mathrm{A}}_{pn}^{-}\right)^{2}\left(\mathrm{B}_{n}^{\left(mp\right)}\mathrm{B}_{n}^{\left(pr\right)}\right)^{\nicefrac{1}{2}}\left(\tau-\tau_{0}\right)$.
The initial waves evolve as

\begin{eqnarray}
\mathrm{A}_{mn}\left(\tau\right) & = & \hat{\mathrm{A}}_{mn}\mathrm{cn}\left(\mathrm{\chi_{n}}|\mathrm{m}_{n}^{-1}\right),\\
\mathrm{A}_{pn}^{-}\left(\tau\right) & = & \hat{\mathrm{A}}_{pn}^{-}\mathrm{dn}\left(\chi_{n}|\mathrm{m}_{n}^{-1}\right).\label{eq: Apq m>1}
\end{eqnarray}
Solutions given by (\ref{eq: Ars  m<=00003D1})-(\ref{eq: Apq m>1})
represent a system wherein energy is shifted periodically between
the three waves. The model demonstrates the most salient features
of resonance: wave generation and cyclical amplitude coupling.

\begin{figure}
\begin{centering}
\includegraphics[scale=0.35]{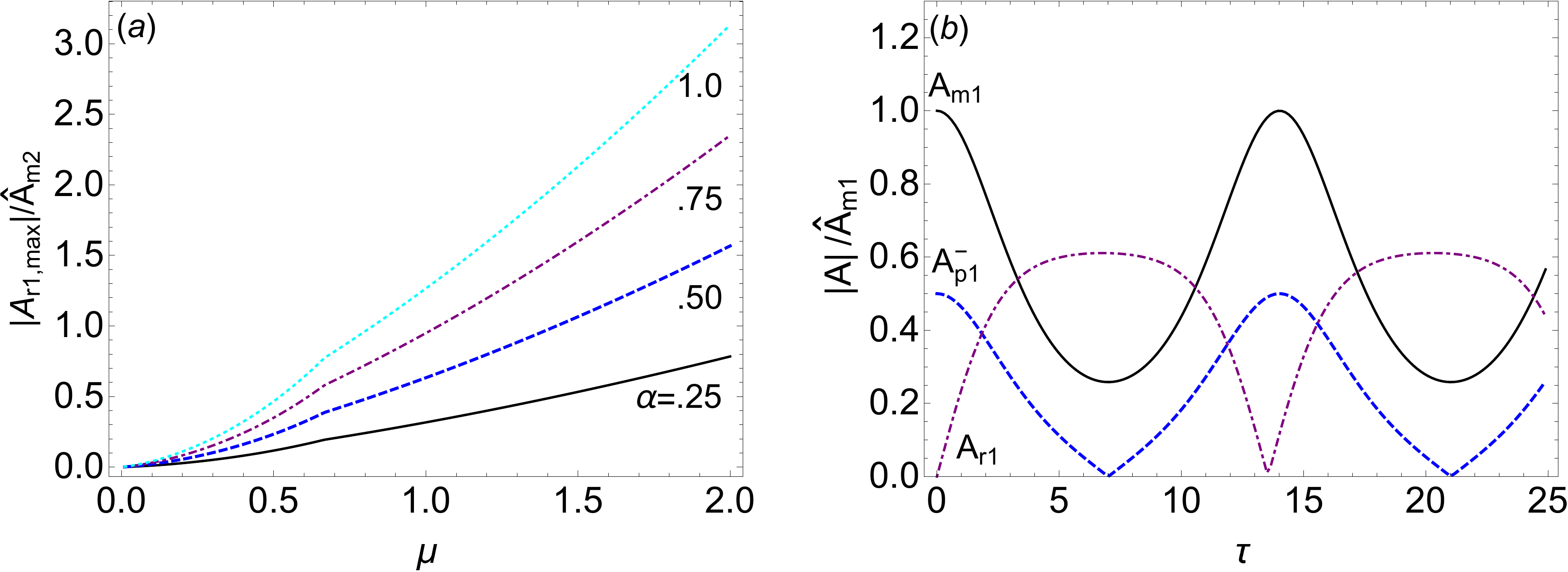}
\par\end{centering}
\caption{{\small{}Type I mode resonance for observing moving with current (a)
Resonant amplitude as function of $\mu$ (ratio of initial harmonics)
for various $\alpha=\hat{\mathrm{A}}_{p1}/\hat{\mathrm{A}}_{m1}$
(ratio of initial amplitudes). Special case of self-interaction, $\mu=1$,
shown for initial condition $\alpha=.5$. Relative maximum resonant
amplitude is $\approx.6$; purple dot-dashed line. (b) Time evolution
of amplitudes: Initial downstream wave $\mathrm{A}_{m1}$; solid black
line, resonates with upstream wave $\mathrm{A}_{p1}^{-}$; blue dashed
line, and generates downstream wave $\mathrm{A}_{r1}$; purple dot-dashed
line. $\mathrm{A}_{r1}$ reaches amplitude $\approx.6$. Since model
parameter $\mathrm{m}_{1}=.993<1$, upstream wave transfers energy
to generated wave more efficiently than $\mathrm{A}_{m1}$ ($\mathrm{A}_{p1}^{-}$
cycles through zero). \label{Case type I modes}}}
\end{figure}

Figure \ref{Case type I modes} shows maximum resonant amplitudes
and coupling for type I mode self-interactions.
 It is shown that increasing $\mu$ and $\alpha$ corresponds to higher
harmonic interactions and greater initial downstream energy respectively.
In the absence of other effects (i.e. wavebreaking and viscous dissipation)
resonant amplitude increases for greater difference between harmonic
interactions and greater initial downstream amplitude (see figure
\ref{Case type I modes}a). In figure \ref{Case type I modes}b resonant
cyclical energy exchange is shown for a self-interaction $\left(\mu=1\right)$
with initial amplitudes $\alpha=\nicefrac{\hat{\mathrm{A}}_{p1}}{\hat{\mathrm{A}}_{m1}}=.5$.
Initial energy on the $\kappa_{m1}$ and $\kappa_{p1}^{-}$ modes
is transferred to the $\kappa_{r1}$ mode which is generated from
zero.

\begin{figure}
\begin{centering}
\includegraphics[scale=0.35]{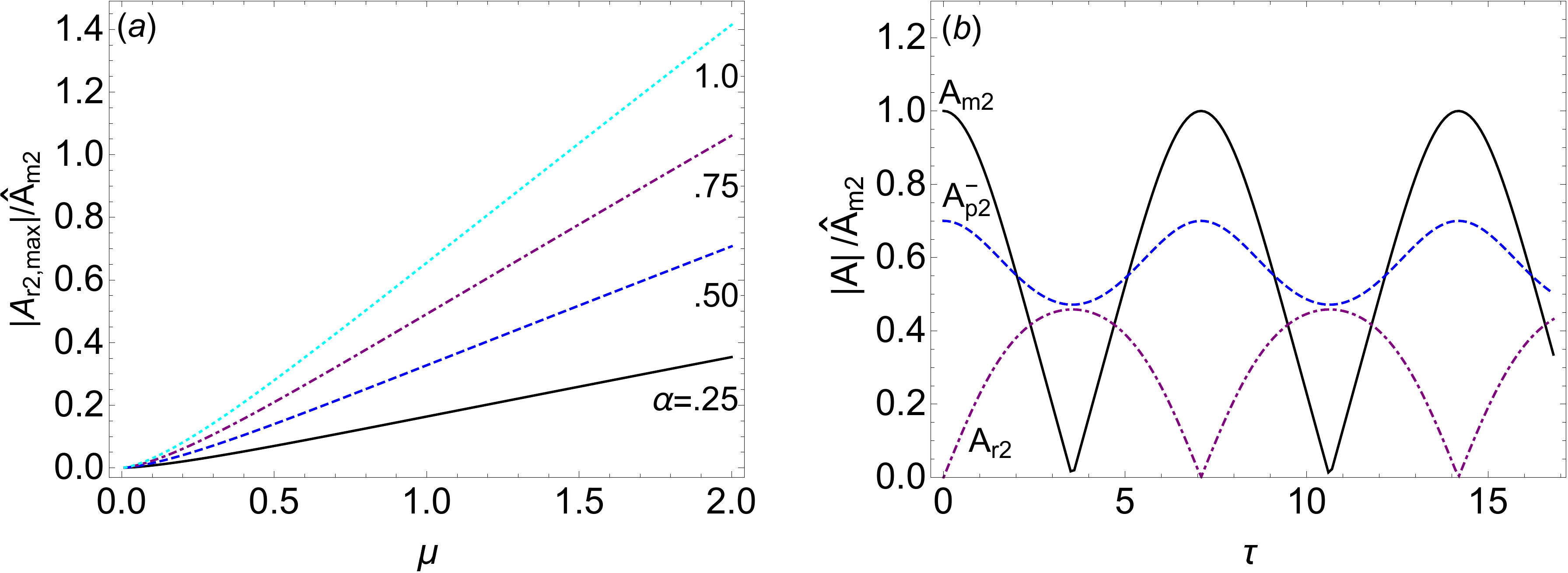}
\par\end{centering}
\caption{{\small{}Type II mode resonance for an observer moving with current
velocity: (a) Resonant amplitude as function of $\mu$ (ratio of initial
harmonics) for various $\alpha=\hat{\mathrm{A}}_{p2}/\hat{\mathrm{A}}_{m2}$
(ratio of initial amplitudes). Consider a self-interaction, $\mu=1$,
and $\alpha=.7$. Relative maximum resonant amplitude is $\approx.42$;
purple-dashed line. (b) Time evolution of amplitudes: Initial downstream
wave $\mathrm{A}_{m2}$; solid black line, resonates with upstream
wave $\mathrm{A}_{p2}^{-}$; blue dashed line, and generates downstream
wave $\mathrm{A}_{r2}$; purple dot-dashed line. $\mathrm{A}_{r2}$
reaches relative amplitude $\approx.42$. Since model parameter $\mathrm{m}_{2}=1.338>1$,
downstream wave transfers energy to generated wave more effectively
than $\mathrm{A}_{p2}^{-}$ ($\mathrm{A}_{m2}$ cycles through zero).\label{Case type II modes}}}
\end{figure}

Figure \ref{Case type II modes} shows maximum resonant amplitudes
and coupling for type II self-interactions.
 In the absence of other effects, maximum amplitudes increase for
higher harmonic interactions (larger $\mu$) and for increasing initial
downstream amplitude (larger $\alpha$) (see figure \ref{Case type II modes}a).
Figure \ref{Case type II modes}b shows cyclical exchange of energy
due to amplitude coupling for a self-interaction and $\alpha=\nicefrac{\hat{\mathrm{A}}_{p2}}{\hat{\mathrm{A}}_{m2}}=.7$.
Initial energy on the $\kappa_{m2}$ and $\kappa_{p2}^{-}$ modes
is transferred to the $\kappa_{r2}$ mode which is generated from
zero. 

Different coupling behaviors may be observed on the wavenumber scales
characteristic of type I and II modes. First, the strength and speed
of the resonant interactions may vary for a given energy distribution
across the wavenumber spectrum (compare figs. \ref{Case type I modes}
and \ref{Case type II modes}). Second, either the initial downstream
wave (for example, figure \ref{Case type I modes}b) or upstream wave
(for example, figure \ref{Case type II modes}b) may cycle through
zero. These behaviors are governed by the parameter $\mathrm{m}_{n}$
which is a function of relative initial amplitudes.

$\mathrm{m}_{n}$ is the effectiveness ratio of the initial waves
in the energy transfer mechanism (see fig. \ref{fig: m dependency}).
When $\mathrm{m}_{n}\leq1$ the upstream mode, $\kappa_{pn}^{-}$,
transfers energy to the resonant wave at a faster rate than $\kappa_{mn}$.
As a result, the $\kappa_{pn}^{-}$ mode cycles through zero. When
$\mathrm{m}_{n}>1$ the roles are reversed. The downstream mode, $\kappa_{mn}$,
transfers energy to the resonant wave at a faster rate than $\kappa_{pn}^{-}$.
In all cases $\mathrm{m}_{n}$ increases with increasing initial downstream
wave amplitude, $\hat{\mathrm{A}}_{mn}$. The regime $\mu<\nicefrac{2}{3}$
is of particular interest, for example, $\mu=.3$ and $\alpha=.25$.
For these values, given the same amplitude ratio on both scales, $\mathrm{m}_{n}>1$
for type I modes but $\mathrm{m}_{n}<1$ for type II modes (compare
figs. \ref{fig: m dependency}a and \ref{fig: m dependency}b). Energy
will be more effectively transferred in different directions on different
scales in wavenumber space. Namely, on the (smaller wavenumber) type
I mode scale energy will be transferred more effectively by the downstream
wave. Whereas, on the (larger wavenumber) type II mode scale, energy
will be transferred more effectively by the upstream wave. For both
mode types greater initial downstream energy (larger $\alpha$) increases
effectiveness of downstream energy transfer to the resonant wave.
Realistically speaking, the energy distribution across wavenumber
scales will not be linear and $\alpha$ may be expected to vary between
these scales.

Resonant behavior was seen to depend on wavenumber mode type and initial
amplitudes. The second wavenumber type is produced by the modified
dispersion relation (see $\mathsection$\ref{subsec:disp relation}).
Initial relative amplitudes would depend on initial energy distribution
as well as current velocity (cf. \citealp{longuet1960changes}). Since
the behavior varies between wavenumber scales (characterized by mode
types), it represents a fundamental source of inhomogeneity. Inhomogeneities
include speed and strength of resonance, and most effective direction
of energy transfer. The resonances and inhomogeneities have no analogy
without current, since the dispersion relation for this case will
be single-valued and isotropic. 

\section{Conclusion \label{sec:Conclusion}}

The main finding of this paper is the existence of new triad resonances
for gravity waves propagating in opposite direction with respect to
a uniform current. The triads are due to multivalued and anisotropic
dispersion. They are a first of their kind for gravity waves in the
sense that they are non-degenerate and occur without inhomogeneity
even in deep water. They are produced by current \textit{vis a vis}
a modified dispersion relation. The most salient features of the resonant
interaction, wave generation and cyclical amplitude coupling, were
shown. Certainly, the assumption of uniform current is an idealization
and capillary and viscous effects will assume importance for large
wavenumbers. Nonetheless, results are expected to be indicative of
a fundamental behavior of wave-current systems under resonance. Analogous
effects may be expected for shearing current which also exhibits multivalued
and anisotropic dispersion relations. 

The linear wave-current dispersion relation was reviewed in $\mathsection$\ref{sec Linear formulation}.
It was shown to be multi-valued and anisotropic. The $existence$
of the new resonances was demonstrated using a geometrical approach
in $\mathsection$\ref{sec: existence of res}. Resonance conditions
for a laboratory observer are derived in $\mathsection$\ref{sec: 4 Res in lab}.
Resonance conditions differ between two wavenumber types (type I and
II modes). Generally speaking, slower currents resonate relatively
higher frequencies (shorter waves) and higher currents resonate lower
frequencies (longer waves) (see Table \ref{tab: Self-interactions-1}).
When the ratio between initial harmonics is known $a$ $priori$,
resonance conditions reduce to universal constants of the non-dimensional
parameter $\mathrm{w}=\omega Ug^{-1}$ (see Table \ref{tab:Universal-constants-1}).
This parameter has been discussed for resonance between waves and
an external forcing \citep[cf.][]{dagan1982free,tyvand2012surface}.
The new resonances suggest the parameter is also an indicator of an
intrinsic energy sharing mechanism within wave-current fields. 

In order to demonstrate physical invariance between inertial observers,
resonance conditions are derived for an observer moving with the velocity
of the current in $\mathsection$\ref{sec: 5 Res in comoving}. The
relevant dispersion relations are not the intrinsic frequencies for
all three waves. The dispersion relation for the wave propagating
against current still ``feels'' an ambient motion in the new frame
of reference due to the preferential direction of the media. Numerical
values for the resonance conditions differ between the laboratory
description and observer moving with current. However, it is in the
latter that analytical solutions take their simplest form. Thus, the
description given by the observer moving with current is preferable
for analysis.

In $\mathsection$\ref{sec: Amplitude coupling and energy exchange},
a three-wave interaction model is adapted for triad resonances on
current. It is used to demonstrate wave generation and energy exchange
between amplitudes. The resonant behavior is dependent on $\mu$ (ratio
of initial harmonics), $\alpha$ (ratio of initial wave amplitudes),
and the permutation $n$ (wavenumber type). These determine the model
parameter $\mathrm{m}_{n}$ (see fig. \ref{fig: m dependency}) which
represents the effectiveness of energy transfer to the resonant wave.
Current changes are known to modify gravity wave amplitudes \citep{longuet1960changes}.
It follows, current magnitude will effect resonant behavior \textit{vis
a vis} changes in the parameters $\alpha$ and $\mathrm{m}$. 

Regimes are seen wherein $\mu$, $\alpha$ and $\mathrm{m}_{n}$ differ
between ``small'' type I and ``large'' type II wavenumber scales
(see fig. \ref{fig: m dependency}). These parameters determine the
dominant direction, speed and strength of energy exchange. Since values
may be different on two coexisting wavenumber scales, different resonance
behaviors may dominate nonlinear energy transfer on different scales
simultaneously (compare figs. \ref{Case type I modes} and \ref{Case type II modes}).
This suggests even uniform current is a fundamental source of spatial
inhomogeneity. 

The results are of both quantitative and qualitative importance. Quantitatively,
their quadratic nonlinearity will dominate other well known higher-order
effects (i.e. quartet interactions). Qualitatively, uniform current
was shown to produce significant inhomogeneity. Moreover, the resonances
represent a fundamental energy transfer mechanism since they preclude
complex 3D effects, topography changes and flow inhomogeneities. Even
without considering current inhomogeneities, the results of this paper
are the basis for consideration of an even richer dynamics which may
be expected for non-collinear wave propagation and capillary effects.\\
\\
Funding: This research was supported by Israel's Ministry of Science
and Technology {[}grant No. 3-12473{]}

\bibliographystyle{plainnat}
\bibliography{Triad_Res_Citations}

\appendix

\section{Symmetry considerations\label{sec:symmetry considerations}}

Intuitions based on the symmetries of the intrinsic frequency dispersion
relation will not hold for water waves propagating on current. The
symmetry of the intrinsic frequency dispersion relation, $\sigma=\sqrt{g\left|k\right|}$,
yields

\begin{eqnarray}
\sigma\left(k\right) & = & \sigma\left(-k\right)\label{eq: frequency symmetry}
\end{eqnarray}
and

\begin{eqnarray}
k\left(\sigma\right) & = & k\left(-\sigma\right),\label{eq: wavenumber symmetry}
\end{eqnarray}
All sign changes yield trivial (non-unique) wavenumber solutions;
waves in opposite directions are defined by negative wavenumbers with
the same magnitude. In practice, these symmetries have lead to simplified
sign conventions in defining resonance interaction closures \citep[i.e.][]{hammack1993resonant}.
However, the symmetries (\ref{eq: wavenumber symmetry}) and (\ref{eq: wavenumber symmetry})
are not generally true for current. An exhaustive list of the symmetry
conditions for wave-current dispersion are not provided here. However,
examples are given which demonstrate that (\ref{eq: frequency symmetry})
and (\ref{eq: wavenumber symmetry}) no longer hold.

For waves following current, $\omega=kU\pm\sqrt{g\left|k\right|}$.
Note, $\omega$ may be understood to be a functional of $\pm\sigma$
because $\pm\sqrt{g\left|k\right|}=\pm\sigma$. It follows immediately

\begin{equation}
\omega\left(k,\sigma\left(k\right)\right)\neq\omega\left(-k,\sigma\left(k\right)\right).
\end{equation}
Thus, (\ref{eq: frequency symmetry}) does not hold for absolute frequencies.

In order to discuss symmetry of the wavenumber solution. Consider
a type I mode solution following current, the solution is

\begin{eqnarray}
k_{m1} & = & \frac{g-2U\omega-\sqrt{g^{2}-4gU\omega}}{2U^{2}}.
\end{eqnarray}
One can see immediately

\begin{eqnarray}
k_{m1}\left(\omega,U\right) & \neq & k_{m1}\left(-\omega,U\right).
\end{eqnarray}
Thus, (\ref{eq: wavenumber symmetry}) may be violated due to the
preferential direction of the media. Note also, the trivial relationship
between wavenumbers for intrinsic frequencies, $k\left(\sigma\right)=k\left(\sigma\right)$
is also no longer true for absolute frequencies

\begin{eqnarray}
k_{m1}\left(\omega_{m},U\right) & \neq & k_{m1}\left(\omega_{m},-U\right).
\end{eqnarray}

As a result of the complex symmetries for waves on current there are
non-trivial (unique) solutions for waves propagating in both directions
with respect to current. And, the transformation between positive
and negative wavenumber conventions is no longer as trivial as it
is for the ``still'' water problem.

\section{Positive and negative wavenumber conventions\label{sec:Positive-and-negative wavenumbers}}

Assuming only positive $\omega$-frequencies, the relative signs of
wavenumbers propagating in both directions with respect to current
are determined by two possible sign conventions. The first, used in
this paper, considers only positive wavenumbers; wave modes propagating
in opposite directions with respect to current will both be positive
in sign, but are distinguishable as they are functions of opposite
signs of $U$. In the second convention, the sign of $U$ is fixed
and oppositely directed waves are given by opposite signs of wavenumbers.
Mixing signs of $k$ and $U$ in any other manner will yield trivial
(non-unique) solutions.

\subsection{Positive wavenumbers convention}

The positive wavenumber convention is employed in this paper. Signs
are chosen such that all wavenumbers are positive; the unique modes
for waves propagating in opposite directions are distinguished by
opposite signs of $U$ in their argument. The four wavenumbers for
this convention, given in (\ref{eq: 4 wavenumbers dimensional}),
are in explicit form

\begin{eqnarray}
k_{m1} & = & \frac{g+2U\omega_{m}-\sqrt{g\left(g+4U\omega_{m}\right)}}{2U^{2}},\label{eq: km 1 (conv1)}\\
k_{m2} & = & \frac{g+2U\omega_{m}+\sqrt{g\left(g+4U\omega_{m}\right)}}{2U^{2}},\label{eq: km 2 (conv1)}\\
k_{m1}^{-} & = & \frac{g-2U\omega_{m}-\sqrt{g\left(g-4U\omega_{m}\right)}}{2U^{2}},\label{eq: km1- (conv1)}\\
k_{m2}^{-} & = & \frac{g-2U\omega_{m}+\sqrt{g\left(g-4U\omega_{m}\right)}}{2U^{2}}.\label{eq: km2- (conv1)}
\end{eqnarray}
The (-)super-script denotes the wave propagating against current ($-U$
has been imposed its definition). As written above, all four solutions
are functions of positive terms only; $U,\omega,g>0$, and when all
solutions are purely real they will also all be positive by definition.

\subsection{Negative and positive wavenumber convention}

A convention which permits for negative wavenumbers is also possible.
In this convention, waves propagating in different directions with
respect to current will have oppositely signed wavenumbers, but are
functions of the same sign of $U$. The wave-current dispersion relation
may be written for this convention as

\begin{eqnarray}
D\left(\omega,k,U\right)=\omega^{2}-2k\left|U\right|\omega+k{}^{2}\left|U\right|^{2}-g\left|k\right| & = & 0.\label{eq: disp rel poly form-1}
\end{eqnarray}
The positive and negative wavenumbers follow immediately from (\ref{eq: disp rel poly form-1}).
Solving (\ref{eq: disp rel poly form-1}) yields,

\begin{eqnarray}
k_{m1} & = & \frac{g+2U\omega-\sqrt{g^{2}+4gU\omega}}{2U^{2}},\label{eq: km1 (conv 2)}\\
k_{m2} & = & \frac{g+2U\omega+\sqrt{g^{2}+4gU\omega}}{2U^{2}},\label{eq: km2 (conv 2)}\\
k_{m1}^{-} & = & \frac{-g+2U\omega+\sqrt{g^{2}-4gU\omega}}{2U^{2}},\label{eq: km1- (conv 2)}\\
k_{m2}^{-} & = & \frac{-g+2U\omega-\sqrt{g^{2}-4gU\omega}}{2U^{2}}.\label{eq: km2- (conv 2)}
\end{eqnarray}
The absolute value of $U$ is dropped but implied. By definition all
wavenumbers are written in terms of positive quantities; $U,\omega,g>0$.
And when all solutions are purely two wavenumbers will be positive
and two negative $by$ $definition$. The (-)superscript denotes the
negative solutions (i.e. waves propagating against current). 

Using these nontrivial positive and negative wavenumbers, resonance
conditions may be recovered. However, with this choice of sign convention,
recovering the triad requires a change in convention in the wave phase
definition. Consider type I modes as an example, the triad resonance
can be closed for the following phasing between nonlinear and linear
terms

\begin{eqnarray}
e^{i\left(k_{m}\left(\omega_{m},\left|U\right|\right)x-\omega_{m}t\right)}e^{-i\left(k_{p}^{-}\left(\omega_{p},\left|U\right|\right)x+\omega_{p}t\right)} & \propto & e^{i\left(k_{r1}\left(\omega_{r},\left|U\right|\right)x-\omega_{r}t\right)}.\label{eq: exotic phasing}
\end{eqnarray}
The resonance closure is

\begin{eqnarray}
\omega_{m}+\omega_{p} & = & \omega_{r},\label{eq: omega  clos exotic}\\
k_{m1}+\left(-k_{p1}^{-}\right) & = & k_{r1}.\label{eq: k close exotic}
\end{eqnarray}
$k_{p1}^{-}$ is negative by definition. In order to avoid ambiguity,
one may take the absolute value of $k_{p1}^{-}$ and reimpose the
negatives in (\ref{eq: k close exotic}). It follows,

\begin{eqnarray}
\omega_{m}+\omega_{p} & = & \omega_{r},\\
k_{m1}+\left(-\left(-\left|k_{p1}^{-}\right|\right)\right) & = & k_{r1}.
\end{eqnarray}
The double negatives yield,

\begin{eqnarray}
\omega_{m}+\omega_{p} & = & \omega_{r1},\label{eq: conv 2}\\
k_{m1}+\left|k_{p1}^{-}\right| & = & k_{r1}.\label{eq: conv 2-1}
\end{eqnarray}
Substitute definitions (\ref{eq: km1 (conv 2)}) and (\ref{eq: km1- (conv 2)})
into (\ref{eq: conv 2}) and (\ref{eq: conv 2-1}) for a self-interaction,
$\omega_{m}=\omega_{p}=\omega$ and $\omega_{r}=2\omega_{m}$. Resonance
is closed when

\begin{equation}
\mathrm{w}=\frac{\omega U}{g}=.178.
\end{equation}
Repeating for type II modes yields the condition $\mathrm{w}=.235$.
These are consistent with the results of $\mathsection$\ref{sec: 4 Res in lab}.

\subsection{Equivalence of conventions}

Denote the wavenumber functions for the second convention, (\ref{eq: km1 (conv 2)})-(\ref{eq: km2- (conv 2)}),
by {*}superscripts. Comparing with (\ref{eq: km 1 (conv1)})-(\ref{eq: km2- (conv1)})
recovers the symmetry rules

\begin{eqnarray}
k_{m1} & = & k_{m1}^{*},\\
k_{m2} & = & k_{m2}^{*},\\
-k_{m1}^{-} & = & k_{m1}^{-*},\\
-k_{m2}^{-} & = & k_{m2}^{-*}.
\end{eqnarray}
The two conventions differ only by a trivial sign change for the waves
propagating against current.

\section{Doppler shift of the resonance closure for waves propagating in opposite
directions on current \label{sec: matrix arguments}}

For a single plane wave, given by a laboratory description, application
of a Doppler shift yields an intrinsic frequency and $vice$ $versa$.
It does not follow that uniform application of a Doppler shift to
a triad resonance closure will always reproduce the resonance closure
in terms of intrinsic frequencies. In particular, this assumption
is not true for a triad resonance between wave modes propagating in
opposite directions with respect to a current (see laboratory description
in $\mathsection$\ref{sec: 4 Res in lab}). This is demonstrated
in the following derivation..

The frequency closures for a triad resonance between absolute and
intrinsic frequencies are

\begin{eqnarray}
\omega_{m}+\omega_{p}^{-} & = & \omega_{r},\label{eq: omega res closure appendix}\\
\sigma_{m}+\sigma_{p} & = & \sigma_{r},\label{eq: sigma}
\end{eqnarray}
respectively. A positive wavenumber convention is employed (see Appendix
\ref{sec:Positive-and-negative wavenumbers} for a discussion of sign
conventions), the ($-$)super-script denotes the wave mode propagating
against current.

As a first step, one can inspect the three independent plane waves
in the same laboratory reference frame. Introducing current into the
velocity potential yields $\left(\omega^{\mp}\right)\pm kU=\sigma$.
Accordingly, the components of (\ref{eq: omega res closure appendix})
and (\ref{eq: sigma}) are related by

\begin{eqnarray}
\left[\begin{array}{c}
\omega_{m}\\
\omega_{p}\\
\omega_{r}
\end{array}\right] & = & \left[\begin{array}{c}
\sigma_{m}\\
\sigma_{p}\\
\sigma_{r}
\end{array}\right]+\left[\begin{array}{ccc}
1 & 0 & 0\\
0 & -1 & 0\\
0 & 0 & 1
\end{array}\right]U\left[\begin{array}{c}
k_{m}\\
k_{p}^{-}\\
k_{r}
\end{array}\right].\label{eq: omega def mat form}
\end{eqnarray}
Equation (\ref{eq: omega def mat form}) is the laboratory description
for the three waves on the same current. Since a positive wavenumber
convention is employed, $-U$ is required to capture the wave propagating
against current (for a negative wavenumber convention, all values
of $U$ are positive but the wavenumber is negative and similar results
follow).

The second step is to apply the same Doppler shift moving with the
velocity $U$ to each frequency component in (\ref{eq: omega def mat form}).
This yields

\begin{gather}
\left[\begin{array}{c}
\sigma_{m}\\
\sigma_{p}\\
\sigma_{r}
\end{array}\right]+\left(\left[\begin{array}{ccc}
1 & 0 & 0\\
0 & -1 & 0\\
0 & 0 & 1
\end{array}\right]-\left[\begin{array}{ccc}
1 & 0 & 0\\
0 & 1 & 0\\
0 & 0 & 1
\end{array}\right]\right)U\left[\begin{array}{c}
k_{m}\\
k_{p}^{-}\\
k_{r}
\end{array}\right]\nonumber \\
\qquad\qquad\qquad=\left[\begin{array}{c}
\sigma_{m}\\
\sigma_{p}\\
\sigma_{r}
\end{array}\right]+\left[\begin{array}{ccc}
0 & 0 & 0\\
0 & -2 & 0\\
0 & 0 & 0
\end{array}\right]U\left[\begin{array}{c}
k_{m}\\
k_{p}^{-}\\
k_{r}
\end{array}\right]\neq\left[\begin{array}{c}
\sigma_{m}\\
\sigma_{p}\\
\sigma_{r}
\end{array}\right].\label{eq: apply gal shift}
\end{gather}
The intrinsic frequencies of all three waves are $not$ recovered.
In (\ref{eq: omega res closure appendix}), the doppler-shifted frequency
of the wave propagating against current is $\sigma_{p}-2Uk_{p}$.
This wave feels a drag $\propto2U$ due to the preferential direction
of the media. This highlights a difficulty in treating a multi-wave
systems on current, since the doppler shift cannot be used to cancel
the current effect for components propagating in opposite directions
with respect to current.

In order to avoid conflating this description with the intrinsic frequency
description, one may introduce a $v$-frequency definition,

\begin{eqnarray}
\left[\begin{array}{c}
v_{m}\\
v_{p}\\
v_{r}
\end{array}\right] & = & \left[\begin{array}{c}
\sigma_{m}\\
\sigma_{p}\\
\sigma_{r}
\end{array}\right]+\left[\begin{array}{ccc}
0 & 0 & 0\\
0 & -2 & 0\\
0 & 0 & 0
\end{array}\right]U\left[\begin{array}{c}
k_{m}\\
k_{p}^{-}\\
k_{r}
\end{array}\right].\label{eq: v res clsoure}
\end{eqnarray}
$v$-frequencies are defined for an observer who moves at the current
velocity, according to the Doppler shift applied in (\ref{eq: apply gal shift}),
and accounts for the preferential direction of the media. The $\upsilon$-description
is not exactly equivalent to three intrinsic frequencies. The dependency
of (\ref{eq: v res clsoure}) on intrinsic frequencies is eliminated
by introducing $local$ definitions for intrinsic frequencies: $\sigma_{i}=\sqrt{gk_{i}}$.
This yields

\begin{eqnarray}
\left[\begin{array}{c}
v_{m}\\
v_{p}\\
v_{r}
\end{array}\right] & = & \left[\begin{array}{c}
\sqrt{gk_{m}}\\
\sqrt{gk_{p}^{-}}\\
\sqrt{gk_{r}}
\end{array}\right]+\left[\begin{array}{ccc}
0 & 0 & 0\\
0 & -2 & 0\\
0 & 0 & 0
\end{array}\right]U\left[\begin{array}{c}
k_{m}\\
k_{p}^{-}\\
k_{r}
\end{array}\right].\label{eq: intrins clos}
\end{eqnarray}
Rearranging (\ref{eq: intrins clos}) yields the following system
of equations

\begin{eqnarray}
v_{m}^{2}-gk_{m} & = & 0,\\
\left(v_{p}^{-}+2k_{p}^{-}U\right)^{2}-gk_{p}^{-} & = & 0,\\
v_{r}^{2}-gk_{r} & = & 0.
\end{eqnarray}
These are the dispersion relations which follow from (\ref{eq: Determinant-1})
in $\mathsection$\ref{sec: 5 Res in comoving}. In $\mathsection$\ref{sec: 5 Res in comoving},
triad resonances were found between $v$-frequencies.

\section{Nonlinear terms and secular growth \label{sec:Nonlinear-terms-secularitiy}}

The nonlinear operator is

\begin{eqnarray}
\mathscr{P}^{2}\left(\tilde{\phi},\tilde{\phi}\right) & = & \frac{1}{g^{2}}\left(-\left(\frac{\partial^{2}\phi}{\partial t^{2}}+2U\frac{\partial^{2}\phi}{\partial x\partial t}+U^{2}\frac{\partial^{2}\phi}{\partial x^{2}}-g\frac{\partial\phi}{\partial z}\right)\left(U\frac{\partial^{2}\phi}{\partial x\partial z}+\frac{\partial^{2}\phi}{\partial z\partial t}\right)\right.\nonumber \\
 &  & -\frac{\partial\phi}{\partial t}\left(U^{2}\frac{\partial^{3}\phi}{\partial x^{2}\partial z}+2U\frac{\partial^{3}\phi}{\partial x\partial z\partial t}+\frac{\partial^{3}\phi}{\partial t^{2}\partial z}+g\frac{\partial^{2}\phi}{\partial z^{2}}\right)+\frac{\partial\phi}{\partial x}\left(2g\frac{\partial^{2}\phi}{\partial x\partial t}\right.\nonumber \\
 &  & \left.-U\left(-2g\frac{\partial^{2}\phi}{\partial x^{2}}+g\frac{\partial^{2}\phi}{\partial z^{2}}+U^{2}\frac{\partial^{3}\phi}{\partial x^{2}\partial z}+2U\frac{\partial^{3}\phi}{\partial x\partial z\partial t}+\frac{\partial^{3}\phi}{\partial t^{2}\partial z}\right)\right).\label{eq: nonlinear terms-1}
\end{eqnarray}
Secular growth in quadratic nonlinear interactions due to resonant
phase matching is seen using the approach of \citet{phillips1957generation,benney1962non}.
One must define a kinematic equation governing the ``slow'' generation
of a resonant wave $b_{rs}$ through the nonlinear interaction of
$b_{mn}$ and $b_{pq}^{-}$. Equations (\ref{eq:linear operator-1}),
(\ref{eq: nonlinear terms-1}) and (\ref{eq: VP three waves}) yield
the nonlinear governing equation of the form given by (\ref{eq: Operator combined form}).

Among linear terms, retain only those involving $b_{rs}$ (the generated
wave). Among nonlinear terms, retain only those $\propto e^{i\left(\left(k_{mn}+k_{pq}^{-}\right)x-\left(\omega_{m}+\omega_{p}\right)t\right)}$.
Let $b_{rs}$ be a function of a ``slow'' time $\tau=\varepsilon t$
(i.e. $b_{rs}=b_{rs}\left(\tau\right)$). According to Resonant Interaction
Theory, non-resonant terms may be assumed to remain of negligible
order. Upon rearranging, growth of the resonant wave in slow time
may be written in the form

\begin{eqnarray}
\frac{db_{rs}}{d\tau} & = & W_{mn,pq,rs}b_{mn}b_{pq}^{-}e^{i\left(\left(k_{mn}+k_{pq}^{-}-k_{rs}\right)x-\left(\omega_{m}+\omega_{p}-\omega_{rs}\right)t\right)}.\label{eq: time evol system A}
\end{eqnarray}
Wherein,

\begin{eqnarray}
W_{mn,pq,rs} & = & \frac{1}{\left(gk_{rs}-(\text{\ensuremath{\omega}}_{r}-k_{rs}U)^{2}\right)}\left(\left(k_{mn}+k_{pq}\right)\left(k_{mn}U+\omega_{mn}\right)\left(k_{pq}U+\omega_{pq}\right)^{2}\right.\nonumber \\
 &  & -gk_{pq}\left(k_{mn}^{2}U-k_{pq}\omega_{mn}-k_{mn}\left(k_{pq}U+\omega_{mn}+2\omega_{pq}\right)\right).\label{eq: Wdef}
\end{eqnarray}
Equation (\ref{eq: Wdef}) is the quadratic nonlinear interaction
coefficient produced by the velocity potential (\ref{eq: VP three waves}).
As expected, the dispersion relation appears in the denominator and
a standard perturbation will fail at exact resonance. It is reassuring
the bound wave solution given in \citet{kadri2013generation} may
be recovered by noting $b\propto ig\left(2\omega\right)^{-1}a$ and
letting $U=0$.

Equation (\ref{eq: time evol system A}) and (\ref{eq: Wdef}) suggests
resonance causes secular growth of nonlinear terms over slow time.
Secular growth is nonphysical. However, equation (\ref{eq: time evol system A})
may be understood to represent an initial growth rate of $b_{rs}$
from zero \citep{phillips1957generation,phillips1960dynamics}. In
order to advance the model to a greater level of realism, all three
amplitudes may be permitted to be functions of the slow time. Amplitude
coupling is then seen to arrest the secular growth \citep[cf.][]{benney1962non,mcgoldrick1965resonant}.
Nonetheless, if the interaction time is sufficiently long, nonlinear
terms may still grow to first-order magnitude (see $\mathsection$\ref{sec: Amplitude coupling and energy exchange}).

\section{Exact solution for a laboratory viewer\label{sec:Appendix exact solutions}}

For a laboratory viewer, the frequency condition for resonance is
(see $\mathsection$\ref{sec: 4 Res in lab})

\begin{eqnarray}
4\mu\mathrm{w}_{mn}+\sqrt{1-4\mu\mathrm{w}_{mn}}+\sqrt{4\mathrm{w}_{mn}+1} & = & \sqrt{4(\mu+1)\mathrm{w}_{mn}+1}+1.\label{eq: gen closure-1}
\end{eqnarray}
An exact solution is given by

\begin{eqnarray}
\mathrm{w}_{mn} & = & -\frac{\sqrt{\beta}}{2}+\frac{1}{2\mu^{2}}+\frac{1}{2\mu}-\left(-1\right)^{n}\frac{\sqrt{\alpha}}{2},\qquad n=1,2.
\end{eqnarray}
Wherein,

{\small{}
\begin{eqnarray}
\alpha & = & \frac{1}{4\sqrt{\beta}}\left(\frac{18}{\mu^{3}}+\frac{54}{\mu^{4}}+\frac{56}{\mu^{5}}+\frac{16}{\mu^{6}}\right)-\frac{4\sqrt[3]{2}}{3\gamma}\left(\frac{4}{\mu^{2}}+\frac{16}{\mu}+22+12\mu+3\mu^{2}\right)\nonumber \\
 &  & +\frac{1}{\mu^{2}}+\frac{5}{3\mu^{3}}+\frac{4}{3\mu^{4}}-\frac{\gamma}{48\sqrt[3]{2}\mu^{6}},\\
\beta & = & \frac{4\sqrt[3]{2}}{3\gamma}\left(\frac{4}{\mu^{2}}+\frac{16}{\mu}+22+12\mu+3\mu^{2}\right)-\frac{1}{3\mu^{3}}+\frac{1}{3\mu^{4}}+\frac{\gamma}{48\sqrt[3]{2}\mu^{6}},\\
\gamma & = & \left(8192\mu^{6}+49152\mu^{7}+116736\mu^{8}+139264\mu^{9}+89856\mu^{10}\right.\nonumber \\
 &  & \left.+32256\mu^{11}+6912\mu^{12}+\delta\right)^{\frac{1}{3}},\\
\delta & = & \left(28311552\mu^{18}+169869312\mu^{19}+401670144\mu^{20}+474218496\mu^{21}\right.\nonumber \\
 &  & \left.+300810240\mu^{22}+106168320\mu^{23}+19464192\mu^{24}\right)^{\frac{1}{2}}.
\end{eqnarray}
}For a self-interaction $\left(\mu=1\right)$, this reduces to

\begin{eqnarray}
\mathrm{w}_{mn} & = & \frac{1}{12}\left(12+\left(-1\right)^{n}3^{2/3}\sqrt{96\sqrt{\frac{1}{\aleph}}-\aleph}-3^{2/3}\sqrt{\aleph}\right).
\end{eqnarray}
Wherein,

\begin{eqnarray*}
\aleph & = & 38\sqrt[3]{\frac{3}{144+\sqrt{159}}}+2\sqrt[3]{144+\sqrt{159}}.
\end{eqnarray*}

\end{document}